        \documentstyle[twoside]{article}

        \catcode`\@=11
        \long\def\@makefntext#1{ 
        \protect\noindent \hbox to 3.2pt {\hskip-.9pt
        $^{{\eightrm\@thefnmark}}$\hfil}#1\hfill} 

        \def\thefootnote{\fnsymbol{footnote}}
         \def\@makefnmark{\hbox to 0pt{$^{\@thefnmark}$\hss}}  

        \def\ps@myheadings{\let\@mkboth\@gobbletwo
        \def\@oddhead{\hbox{} 
        \rightmark\hfil\eightrm\thepage}
        \def\@oddfoot{}\def\@evenhead{\eightrm\thepage\hfil 
        \leftmark\hbox{}}\def\@evenfoot{}
        \def\sectionmark##1{}\def\subsectionmark##1{}}

        \oddsidemargin=\evensidemargin
        \renewcommand{\thefootnote}{\fnsymbol{footnote}}

\newcounter{sectionc}\newcounter{subsectionc}\newcounter{subsubsectionc}
        \renewcommand{\section}[1] {\vspace{12pt}\addtocounter{sectionc}{1}
        \setcounter{subsectionc}{0}\setcounter{subsubsectionc}{0}\noindent
                {\tenbf\thesectionc. #1}\par\vspace{5pt}}
        \renewcommand{\subsection}[1]
{\vspace{12pt}\addtocounter{subsectionc}{1}
                \setcounter{subsubsectionc}{0}\noindent
                {\bf\thesectionc.\thesubsectionc. {\kern1pt \bfit
#1}}\par\vspace{5pt}}
        \renewcommand{\subsubsection}[1]
{\vspace{12pt}\addtocounter{subsubsectionc}{1}

\noindent{\tenrm\thesectionc.\thesubsectionc.\thesubsubsectionc.
                {\kern1pt \tenit #1}}\par\vspace{5pt}}
        \newcommand{\nonumsection}[1] {\vspace{12pt}\noindent{\tenbf #1}
                \par\vspace{5pt}}
        \newcounter{appendixc}
        \newcounter{subappendixc}[appendixc]
        \newcounter{subsubappendixc}[subappendixc]
        \renewcommand{\thesubappendixc}{\Alph{appendixc}.\arabic{subappendixc}}
        \renewcommand{\thesubsubappendixc}

{\Alph{appendixc}.\arabic{subappendixc}.\arabic{subsubappendixc}}

        \renewcommand{\appendix}[1] {\vspace{12pt}
                \refstepcounter{appendixc}
                \setcounter{figure}{0}
                \setcounter{table}{0}
                \setcounter{lemma}{0}
                \setcounter{theorem}{0}
                \setcounter{corollary}{0}
                \setcounter{definition}{0}
                \setcounter{equation}{0}
                \renewcommand{\thefigure}{\Alph{appendixc}.\arabic{figure}}
                \renewcommand{\thetable}{\Alph{appendixc}.\arabic{table}}
                \renewcommand{\theappendixc}{\Alph{appendixc}}
                \renewcommand{\thelemma}{\Alph{appendixc}.\arabic{lemma}}
                \renewcommand{\thetheorem}{\Alph{appendixc}.\arabic{theorem}}

\renewcommand{\thedefinition}{\Alph{appendixc}.\arabic{definition}}

\renewcommand{\thecorollary}{\Alph{appendixc}.\arabic{corollary}}
                \renewcommand{\theequation}{\Alph{appendixc}.\arabic{equation}}
                \noindent{\tenbf Appendix \theappendixc #1}\par\vspace{5pt}}
        \newcommand{\subappendix}[1] {\vspace{12pt}
                \refstepcounter{subappendixc}
                \noindent{\bf Appendix \thesubappendixc. {\kern1pt \bfit #1}}
                \par\vspace{5pt}}
        \newcommand{\subsubappendix}[1] {\vspace{12pt}
                \refstepcounter{subsubappendixc}
                \noindent{\rm Appendix \thesubsubappendixc. {\kern1pt \tenit
#1}}
                \par\vspace{5pt}}

        \newcommand{\textlineskip}{\baselineskip=13pt}
        \newcommand{\smalllineskip}{\baselineskip=10pt}

        \def\eightcirc{
        \begin{picture}(0,0)
        \put(4.4,1.8){\circle{6.5}}
        \end{picture}}
        \def\eightcopyright{\eightcirc\kern2.7pt\hbox{\eightrm c}}
        \newcommand{\copyrightheading}[1]
                {\vspace*{-2.5cm}\smalllineskip{\flushleft
                {\eightrm Modern Physics Letters A, #1}\\
                {\eightrm $\eightcopyright$\, World Scientific Publishing
                 Company}\\
                 }}

        \newcommand{\publisher}[2]{{\begin{center}\eightrm\smalllineskip
                Received #1\\
                Revised #2
                \end{center}
                }}

        \def\abstracts#1#2#3{{
                \centering{\begin{minipage}{4.5in}\baselineskip=10pt\eightrm
                \centerline{ABSTRACT}
                \parindent=0pt #1\par
                \parindent=15pt #2\par
                \parindent=15pt #3
                \end{minipage} }\par}}

        \renewenvironment{thebibliography}[1]                   
                {\ninerm
                 \baselineskip=11pt                             
                 \begin{list}{\arabic{enumi}.}
                {\usecounter{enumi}\setlength{\parsep}{0pt}
                 \setlength{\leftmargin 17pt}{\rightmargin 0pt} 
                 \setlength{\itemsep}{0pt} \settowidth          
                {\labelwidth}{#1.}\sloppy}}{\end{list}}

        \newcounter{itemlistc}
        \newcounter{romanlistc}
        \newcounter{alphlistc}
        \newcounter{arabiclistc}
        \newenvironment{itemlist}
                {\setcounter{itemlistc}{0}
                 \begin{list}{$\bullet$}
                {\usecounter{itemlistc}
                 \setlength{\parsep}{0pt}
                 \setlength{\itemsep}{0pt}}}{\end{list}}

        \newcommand{\fcaption}[1]{
                \refstepcounter{figure}
                \setbox\@tempboxa = \hbox{\eightrm Fig.~\thefigure. #1}
                \ifdim \wd\@tempboxa > 5in
                   {\begin{center}
                \parbox{5in}{\eightrm \smalllineskip Fig.~\thefigure. #1 }
                    \end{center}}
                \else
                     {\begin{center}
                     {\eightrm Fig.~\thefigure. #1}
                      \end{center}}
                \fi}

        \newcommand{\tcaption}[1]{
                \refstepcounter{table}
                \setbox\@tempboxa = \hbox{\eightrm Table~\thetable. #1}
                \ifdim \wd\@tempboxa > 5in
                   {\begin{center}
                \parbox{5in}{\eightrm\smalllineskip Table~\thetable. #1 }
                    \end{center}}
                \else
                     {\begin{center}
                     {\eightrm Table~\thetable. #1}
                      \end{center}}
                \fi}

        \def\@citex[#1]#2{\if@filesw\immediate\write\@auxout    
                {\string\citation{#2}}\fi                       
        \def\@citea{}\@cite{\@for\@citeb:=#2\do                 
                {\@citea\def\@citea{,}\@ifundefined             
                {b@\@citeb}{{\bf ?}\@warning
                {Citation `\@citeb' on page \thepage \space undefined}}
                {\csname b@\@citeb\endcsname}}}{#1}}

        \newif\if@cghi
        \def\cite{\@cghitrue\@ifnextchar [{\@tempswatrue
                \@citex}{\@tempswafalse\@citex[]}}
        \def\citelow{\@cghifalse\@ifnextchar [{\@tempswatrue
                \@citex}{\@tempswafalse\@citex[]}}
        \def\@cite#1#2{{$\null^{#1}$\if@tempswa\typeout
                {IJCGA warning: optional citation argument
                ignored: `#2'} \fi}}

        \def\pmb#1{\setbox0=\hbox{#1}
                \kern-.025em\copy0\kern-\wd0
                \kern.05em\copy0\kern-\wd0
                \kern-.025em\raise.0433em\box0}

        \def\fnt#1#2{\footnotetext{\kern-.3em
                {$^{\mbox{\scriptsize #1}}$}{#2}}}

        \def\fpage#1{\begingroup
        \voffset=.3in
        \thispagestyle{empty}\begin{table}[b]\centerline{\footnotesize #1}
                \end{table}\endgroup}

        \def\runninghead#1#2{\pagestyle{myheadings}
        \markboth{{\eightit{\quad #1}}\hfill}{\hfill{\eightit{#2\quad}}}}

        \font\tenbf=cmbx10
        \font\tenit=cmti10
        \font\tenit=cmti10
        \font\bfit=cmbxti10 at 10pt
         1
         1
         1
        
        \font\ninerm=cmr9

        \font\eightrm=cmr8
        \font\eightit=cmti8

        \def\qed{\hbox{${\vcenter{\vbox{                          
           \hrule height 0.4pt\hbox{\vrule width 0.4pt height 6pt
           \kern5pt\vrule width 0.4pt}\hrule height 0.4pt}}}$}}

        \runninghead{Recent Developments in String Field Theory in
        D=2 $\ldots$} {Recent Developments in String Field Theory in
        D=2 $\ldots$}

\begin{document}
        \normalsize\textlineskip
        {\thispagestyle{empty}
        \setcounter{page}{1}

        \renewcommand{\thefootnote}{\fnsymbol{footnote}} 

        \copyrightheading{Vol. 0, No. 0 (1992) 000--000}

        \vspace*{0.88truein}

        \fpage{1}
        \centerline{\bf RECENT DEVELOPMENTS IN D=2 STRING FIELD THEORY}
        \vspace{0.37truein}
        \centerline{\footnotesize MICHIO KAKU}
        \vspace*{0.015truein}
        \centerline{\footnotesize\it Physics Dept, City College of the
        City University of New York}
        \baselineskip=10pt
        \centerline{\footnotesize\it New York, N.Y., 10031, USA}
        \vspace{0.225truein}
        \publisher{(received date)}{(revised date)}
        \vspace*{0.21truein}
        \abstracts{\noindent
        We review the recent developments in constructing
        string field theory in two-dimensions. We
        analyze the bewildering number of string field theories
that have been proposed,
all of which correctly reproduce the correlation functions
of two-dimensional string theory. These include:
        \begin {itemlist}
        \item  free fermion field theory
        \item collective string field theory
        \item temporal gauge string field theory
        \item non-polynomial string field theory
        \end {itemlist}
        We will analyze discrete states, the $w(\infty)$
        symmetry, and correlation functions in terms of
        these different string field theories. We will also
comment on the relationship between these various
field theories, which is still not well understood.}{}{}

        \vspace*{-3pt}\textlineskip
        \section{Introduction}
        At present, string theory$^1$ gives us the best hope for a
        unified theory of
        all known interactions, including gravity.
        However, the frustrating aspect of
        string theory is that millions of classical solutions are now
        known of the string equations, perhaps corresponding to the
        set of all possible conformal field theories$^2$.
        Which one, if any, describes the physical world?

        There are indications that the theory is not Borel summable,
        meaning that these perturbative vacua are probably unstable,
        and that the true vacuum of the theory must be found
        non-perturbatively.
        Thus, the principle problem of string theory is
        to determine its true, non-perturbative vacuum.
And the most conservative method of
        formulating a non-perturbative framework for string
        theory is through a second quantized
        string field theory$^3$.

        A first quantized theory is essentially a single-particle
formalism, where the co-ordinates of a single particle $x _ \mu $ are
       quantized, according to the relations:

        \begin {equation}
        [ x _ \mu , p _ \nu ] = i \delta _ { \mu \nu}
        \end {equation}
        Because the first quantized theory describes the
co-ordinates of a single particle $x _ \mu ( t )$,
it cannot easily describe multi-particle states.  In particular,
        it is difficult to describe the true vacuum of the theory.

        In a first quantized point-particle
formalism, interactions can be introduced,
but only in a rather awkward fashion, by specifying by hand
the set of all possible interacting graphs.
We define the scattering amplitude
        via:

        \begin {equation}
        A ( k _ 1, k _ 2 , ...  ) \sim
        \sum _ { \rm topologies}
        \int D x _ \mu ( t ) \, \exp \left ( i S + \sum _ j i k _ j x _ j
\right)
        \end {equation}
        where we must arbitrarily impose the set of topologies
        (corresponding to the complete set of Feynman diagrams)
        over which
        we perform the functional integration.
        Thus, the topologies of the Feynman diagrams, the
various relative
        weights of the Feynman diagrams, the particular choice
        of which Feynman diagrams to include, etc. are all
        rather arbitrary. In particular, it is difficult to prove
        the unitarity of the theory.

        A second quantized theory, by contrast, is necessarily a
        multi-particle formalism. We begin not with the co-ordinates
of a single particle,
        but with the field variables $\phi$ which can describe
multi-particle states, and define the commutation
        relations to be:

        \begin {equation}
        [ \phi ( x ) , \pi ( y ) ] = i \delta ( x - y )
        \end {equation}
        for equal times.
        In contrast to the first quantized theory, where
the set of Feynman diagrams must be imposed as an additional,
cumbersome
constraint, in a second quantized theory everything is
fixed once we specify the invariant
        Lagrangian. This in turn automatically yields the
        complete topology of Feynman diagrams, their relative weights,
        etc. Unitarity is proven simply by showing that the
        Hamiltonian is hermitian.

        In principle, a second quantized
string field theory should yield the
        true vacuum of string theory. Unfortunately, in practice
        string field theory has fallen far short of this ambitious goal,
        in part because it is defined in a specific background.
        At present, string field theory has only been able to
        reproduce the known results of perturbative string theory,
        and has provided little useful non-perturbative information.

        In recent years, the most promising avenue to approach
        non-perturbative string theory
        has been to explore two-dimensional
        string theory$^4$, which is, remarkably enough, exactly
        soluble. Thus, two-dimensional string theory serves as
        a toy model or a theoretical laboratory in which
        one can obtain valuable intuitive insight into the
        full theory.

        In two-dimensions, in turn,
        there are two ways in which to approach
        string theory:

        1) We can use \lq\lq matrix models," in which we approximate
        the Riemann surfaces found in string theory via
        a discrete triangulation and then take the continuum limit.
        $c = 1$ matrix models
        begin with the one dimensional action:

        \begin {equation}
        L = {1\over 2} \, {\rm Tr } \, \left( \dot{M}^2 - U(M)\right)
        \end {equation}
        where $U(M)$ is the potential, and
        where $M(t)$ is a $N \times N$
        hermitian matrix which is a function of time $t$.
        (The eigenvalues of $M$ will eventually become a
        continuous variable, which corresponds to the
        second dimension of space-time.)
        The great advantage of using matrix models is that they are
        exactly soluble and simple to use.

        However, the disadvantage of matrix models is that
        almost
        all string degrees of freedom have been eliminated, and hence
        the intuitive concept of the
        string, in some sense, has disappeared.
        All we have left is the massless tachyon.
        This means that the rather mysterious
        \lq\lq discrete states"
        and the $w(\infty)$ algebra appear in a rather
        subtle and obscure fashion.

More seriously, matrix models are only defined for $c \leq 1$,
meaning that they cannot yield any realistic information about our
physical world. This is a severe drawback, which has retarded
the development of this approach.

        2) We can formulate two-dimensional string theory via
        Liouville theory$^5$.

        Liouville theory begins with the first quantized action:

        \begin {equation}
\label{eq:louie}
S = { 1 \over 8 \pi }
 \int d^2 \sigma \sqrt{\hat{g}}
        \left[\hat{g}^{ab} \left(\partial _a X^i\partial
_b X^i + \partial _ a\phi
        \partial _b \phi \right) -
 Q \hat{R} \phi + \mu e^{ - { \sqrt 2 }\phi}
        \right]\ .
        \end {equation}
where $\phi$ is the Liouville field, a
remnant of the original world-sheet metric
$g _ { ab }$, and $\mu$ is the cosmological constant
on the world-sheet.
(This conformal action,
where $\phi$ is treated as time,
is written in the Euclidean metric.
We can make the transition to the Minkowski metric
with the substitution $\phi \rightarrow i t$.)

        The advantage of the Liouville approach is that
        it is
        transparently a string theory. All string degrees of
        freedom, including the familiar ghosts, are present.
The tachyon, for example, appears with mass
$ m ^ 2 = { 1 \over 12 } ( 2 - D )$, which vanishes if $D =2$.

When re-expressed as a string field theory, this formalism also
enjoys the advantage that the discrete states and $w(\infty)$
algebra appear naturally. As in gauge theory, we find that
the three-string coupling constants are exactly the structure
constants of $w(\infty)$, and that the discrete states
emerge as solutions of the gauge constraints. In other words,
the discrete states and $w(\infty)$ emerge as byproducts of
the gauge symmetry of string field theory, giving us an intuitive
understanding of why they exist.

The Liouville action has some rather peculiar features.
Because the action is not translation invariant in $\phi$,
we will find
that the light-cone gauge cannot be naively imposed on the theory.
This means that we cannot use the gauge conditions to eliminate all
higher
string degrees of freedom. (This, in turn, helps to
explain why there are discrete states which cannot be eliminated
in the light-cone gauge.) Because translation invariance is broken,
we will also find that the usual conservation of energy is
violated.

        However, the most serious complication of this action is
that it is highly non-linear and, even at the free level,
        almost intractable.
        This is because the classical equation of motion for
        the Liouville field is $\partial ^ 2 \phi \sim
        e ^ { \alpha \phi }$, which is highly non-linear.
        Most results for Liouville theory are formulated
        in the approximation where the cosmological
        constant is taken to be zero: $\mu = 0$. In this limit,
        the theory becomes a free theory, and all correlation
        functions are easily found. These scattering amplitudes
        correspond to what is called \lq\lq bulk" scattering.
        (For non-zero cosmological constant,
there is an impenetrable barrier when the
$e ^ { \alpha \phi}$ term becomes large,
so we have scattering or reflection off a
wall. These scattering amplitudes for non-zero $\mu$
correspond to what is called \lq\lq wall" scattering,
        and are much more difficult to solve.)

        Progress in this area has been painstakingly slow.
        In fact, the explosion of research in matrix models
        helped to re-stimulate research in Liouville theory,
        as surprisingly simple results were found for
        correlation functions which were previously thought,
        using Liouville theory, to be
        intractable.

        In this paper, we will study D = 2 string field
        theory from both perspectives.
        We will first develop matrix models via the
        theory of collective string
        field theory, and then we will discuss the
        string field theory of Liouville theory.

        Our task is complicated by the fact that there are
        a bewildering variety of string field theories in
        two-dimensions, including:

(a) free fermion field theory$^6$

(b) collective field theory of Das and Jevicki$^7$

(c) temporal gauge field theory of Ishibashi and Kawai$^8$

(d) the non-polynomial
        BRST string field theory$^9$ of the author.

(The 2D non-polynomial string field theory is a
modification of the non-polynomial string field
found in 26 dimensions by the author$^{10}$
 and the MIT$^{11}$ and Kyoto groups$^{12}$.)

        Eventually, we hope that all of them can be shown to be nothing
        but different
        gauge choices of the same string field theory.
        For example, it is expected that the BRST Liouville
        string field theory can be reduced to the
        collective string field
        theory when all the BRST trivial states are
        eliminated.
        Because reparametrization invariance allows us to cancel
        two degrees of freedom, and since there are only two degrees of
        freedom in $D =2$, this means that collective field theory
        is defined only in terms of a single tachyon, while
        BRST string field theory includes all redundant
        string states as well as BRST trivial states.

        However, at present the complete
relationship between all of these
        formalisms is largely unknown.

        \section {Matrix Models and Free Fermion Theory}

        In this section, we will first show that matrix models
        can be reduced to string theory in the
        double scaling limit, and that the theory
        is soluble because it is equivalent to a free fermion
        field theory.

        Let us perform a simple counting of the various
        $N$ factors found in the Feynman rules generated
by the action for matrix models.
We find:
\begin {equation}
\left \{ \begin {array} {lll}
        V = {\rm vertices} & \rightarrow & N
        \\
        P = {\rm propagators} & \rightarrow & N ^ { -1 }
        \\
        L = {\rm loops } & \rightarrow & N
        \\
        \end {array}
        \right .
        \end {equation}

          The partition function
        therefore has the form:

        \begin {equation}
        Z ( g )
        =
        \sum _ h N ^ { V - P + L } Z _ h ( g )
        = \sum _ h N ^ { 2 - 2h } Z _ h ( g )
        \end {equation}
This means that each Feynman diagram
        contributes an overall factor of
        $ N ^ { 2 - 2h }$, where $h$ is the number
        of handles or closed loops of the diagram, and
where $\chi = 2 - 2h$ is the Euler number.
Thus, a $1/N$ expansion decomposes the perturbation theory
in terms of topologically equivalent diagrams.
In the limit $N \rightarrow \infty$, only the planar
        diagram, with no loops, survives. This limit is not
        that interesting. More instructive is taking the
        limit as the coupling $g$ approaches some critical
        value, which alters the nature of the limit process.
Then the combination of $N \rightarrow\infty$
with $g \rightarrow g _ c$ gives us a non-trivial limit.

        For example, in string theory
it can be shown  that $Z _ h$ obeys the
        property$^{13}$:

        \begin {equation}
        Z _ h ( g ) \sim f _ h (g _ c - g ) ^ { ( 2 - \Gamma ) \chi / 2 }
        \end {equation}
        where $g _ c$ represents a critical value for the coupling
        constant, $\chi$ is the Euler number of the surface,
 and $\Gamma$ is the critical exponent,
        given by $\Gamma = ( 1 / 12 )
        ( D - 1 - \sqrt { ( D-1)(D-25)} )$.

        Let us define $\kappa  ^ { -1 } =  N ( g - g _ c ) ^ { ( 2 - \Gamma)/ 2
} $.
        Thus, we can define a \lq\lq double scaling limit,"
        where we take the limit $N \rightarrow \infty$,
        $ g \rightarrow g _ c$, and:

        \begin {equation}
        {\rm double \,\, scaling\,\, limit}
        = \left \{
        \begin {array} {lll}
        N & \rightarrow & \infty
        \\
        g & \rightarrow & g _ c
        \\
        \kappa & \rightarrow & {\rm constant}
        \\
        \end {array}
        \right .
        \end {equation}

        Then the partition function becomes:

        \begin {equation}
        Z ( g ) \rightarrow \sum _ h \kappa ^ { 2h - 2 } f _ h
        \end {equation}
        Thus, in the double scaling limit, we expect that
matrix models provide a good description of string theory
in two-dimensions. We should also point out that, in the
double scaling limit, the precise nature of the
potential $U(M)$ washes out, and so the final results are
largely independent of the nature of $U(N)$.

        The next step is to write matrix models as a field theory.
To accomplish this, we will
first write out the Hamiltonian and then
make a change of variables:

                      \begin {eqnarray}
                H & = & -  {  1 \over 2 } \Delta + U
                \nonumber \\
                \Delta & = &
                \sum _ i { \partial ^ 2 \over
                \partial M _ { ii } ^ 2 } +
                 { 1 \over 2 } \sum _ {  i < j }
                { \partial ^ 2 \over \partial \,
                {\rm Re} \, M _ { ij } ^ 2 }
                + { \partial ^ 2 \over \partial
                \, {\rm Im} \, M _ { i j } ^ 2 }
                \nonumber \\
                U & = &
                { 1 \over 2 }\,  {\rm Tr}\, M ^ 2 +
                 { g \over N }\, {\rm Tr} \, M ^ 4
                \end {eqnarray}

Likewise, the functional measure is given by:

              \begin {equation} \prod _ { ij }  d M _ { ij }
                \equiv \prod _ i
                d M _ { ii } \prod _ { i < j }
                d ( {\rm Re}\,  M _ {ij } ) d (
        {\rm Im} \,  M _ { ij } )
                \end {equation}

Our strategy is to rewrite this Hamiltonian and
measure in terms of the eigenvalues
$\lambda _i $ of the matrix $M$, as well as the
angular variables.

We can always decompose the matrix $M(t)$ into:

\begin {equation}
M  (t)=\Omega^\dagger(t) \Lambda(t) \Omega(t)
\end {equation}
where
$\Omega \in SU(N)$ and $\Lambda= \, {\rm diag} \,  (\lambda_1, \lambda_2,
\ldots,
\lambda_N)$.
With this decomposition, the Lagrangian
now depends on the angular part $\Omega$ via the term:

\begin {equation}
{\rm Tr } \,
\dot M ^ 2=\, {\rm Tr} \,\dot\Lambda^2+\, {\rm Tr} \,[\Lambda,
\dot\Omega\Omega^\dagger]^2\ .
\end {equation}

To eliminate this last term, it will be helpful to
decompose it in terms of the generators of $SU(N)$:

\begin {equation}
\dot\Omega\Omega^\dagger
={i\over\sqrt 2}\sum_{i<j}\dot\alpha_{ij}
T_{ij}+\dot\beta_{ij}\tilde T_{ij}+
\sum_{i=1}^{N-1}\dot\alpha_i H_i
\end {equation}
where $H_i$ are the diagonal generators of the Cartan subalgebra.
$T_{ij}$ is defined to be the matrix $M$ such
that $M_{ij}=M_{ji}=1$, and
zero elsewhere;
$\tilde T_{ij}$ is the matrix $M$ such that $M_{ij}=-M_{ji}=-i$, and
all other entries are 0.

The Lagrangian is now written as:

\begin {equation}L=\sum_i\left(
\dot\lambda _ i ^2+U(\lambda _ i ) \right)
+  \sum_{i<j} (\lambda _ i - \lambda _ j )^2(
\dot\alpha_{ij}^2 +\dot\beta_{ij}^2)
\end {equation}

We also note that the
measure of integration, in these
new co-ordinates, looks like
$ {\cal D} \Phi ={\cal D} \Omega\prod_i d\lambda_i \Delta^2(\lambda)$,
where $\Delta(\lambda)$~is the Vandermonde determinant
$\prod_{i<j}(\lambda _i-\lambda_j)$. Because of the non-trivial
Jacobian, the kinetic term for the eigenvalues becomes

\begin {equation}-{1 \over 2 \beta^2 } \sum_{i=1}^N
{1\over\Delta^2(\lambda )}{d\over d \lambda_i}
\Delta^2(\lambda){d\over d \lambda_i}
\end {equation}
(where $\beta$ corresponds to an effective
inverse Planck's constant, which goes as $N$).

This can also be rewritten as:
\begin {equation}-{1 \over 2 \beta^2 \Delta(\lambda )} \sum_i{d^2\over {d
\lambda_i}^2}
\Delta(\lambda)\ .
\end {equation}

Then the total Hamiltonian can be expressed as:

\begin {equation}
H= -{1 \over 2 \beta^2 \Delta(\lambda )} \sum_i{d^2\over {d \lambda_i}^2}
\Delta(\lambda ) + \sum_i U(\lambda_i)
+\sum_{i < j} {\Pi_{ij}^2+\tilde\Pi_{ij}^2
 \over {(\lambda_i - \lambda_j)^2}}
\end {equation}
where $\Pi_{ij}$ and $\tilde\Pi_{ij}$ are
conjugate to $\alpha_{ij}$ and $\beta_{ij}$,
that is,
they are the generators of {\it left} rotations on $\Omega$,
$\Omega \rightarrow A\Omega$.

Fortunately, we only need concern ourselves with the
$SU(N)$
singlet sector of the theory. This is because
the angular terms in the Hamiltonian
are all positive definite. To find the
ground state of the system,
we therefore need only
examine the singlet sector.
Our problem is thus reduced to examining the system
defined by:

\begin {eqnarray}
\left( \sum_{i=1}^N h_i\right)
\Psi (\lambda)& = & E\Psi (\lambda)\
\nonumber \\
h_i & =& -{1 \over 2 \beta^2 } {d^2\over {d \lambda_i}^2}
+U(\lambda_i)
\end {eqnarray}
where $\Psi(\lambda) =\Delta(\lambda)
\chi_{sym}(\lambda)$ is by construction antisymmetric
because of the presence of the Vandermonde determinant
$\Delta$.

Several key points must be made. First, the wave function
$\Psi$, because of the presence of the Vandermonde determinant,
is now fermionic. Second, the theory has now
reduced to a system of uncoupled non-relativistic
fermions acting under a
potential.
This is one way to see why the theory is exactly soluble.
Although the boundary conditions may be a bit involved,
the Hamiltonian is that of a free fermion theory.

The fermionic string field $\Psi$ can  be written as a
linear superposition of an infinite number of states
        $\psi _ i$ with energy $ e _ i $:

        \begin {equation} \Psi ( \lambda , t )   = \sum _ i \alpha _ i \psi _ i
( \lambda )
        e ^ { - i e _ i t }
        \end {equation}

        In this representation, we can easily re-write our original Hamiltonian
        in terms of this string field:

        \begin {equation}
\label {eq:fermion}
H =  \int d \lambda \left [
        { 1 \over 2 \beta ^ 2 }
        { \partial \Psi ^ \dagger \over
        \partial \lambda }
        { \partial \Psi \over
        \partial \lambda }
        +
        U( \lambda ) \Psi ^ \dagger \Psi
        - \mu _ F ( \Psi ^ \dagger \Psi - N)
        \right]
        \end {equation}
        (we adjust the Lagrange multiplier $\mu _F $ to equal the Fermi
        level).

        This Hamiltonian, in turn, can be derived from the following
        action:

        \begin {equation} S = \int d t d \lambda \left [
        i \Psi ^ \dagger \dot \Psi
        -
         { 1 \over 2 \beta ^ 2 }
        { \partial \Psi ^ \dagger \over
        \partial \lambda }
        { \partial \Psi \over \partial \lambda }
        - U( \lambda )
        \Psi ^ \dagger \Psi +
        \mu _ F ( \Psi ^ \dagger \Psi - N ) \right ]
        \end {equation}
        This string field theory action contains all the information
        of the $c=1$ matrix model for the $SU(N)$ singlet sector.
        Notice that the action appears to
        be defined in {\it two} dimensions if we treat $\lambda$ and $t$ as
        two space-time co-ordinates. This is how the Liouville mode appears
        in our formulation.

        We should point out that, in this form, the free fermion action
can yield many interesting features of two-dimensional string theory,
such as the energy states, the scattering amplitudes, etc.
However, instead of developing this formalism any further, let us now
discuss a new formalism, collective string field theory,
and then show the equivalence of free fermion field theory
with collective string field theory.

        \section {Collective Field Theory}

        Next, we wish to develop the formalism of collective field
        theory of Jevicki and Sakita$^{14}$
and derive the collective string field theory action.
        Collective field theory involves a change of variables,
        from an original set of fields to a set of invariant
        ones. Originally, the goal of collective field theory was
        to replace the variable $A _ \mu ^ a$ appearing in gauge
        theory with the Wilson loop, involving an invariant trace
        of a closed loop.
        Although this has not been particular successful, it has
        enjoyed considerable success for two-dimensional string theory.

        We start with a standard
        Hamiltonian defined with variables $q _ i$:

        \begin {equation}
        H = - { 1 \over 2 } \sum _ i ^ N {\partial ^ 2 \over \partial q _ i ^ 2
}
        + V ( q _ i )
        \end {equation}

        Next, we change variables from the $q _ i $
        to a field variable $\phi ( x )$ as follows:

        \begin {equation}
        \phi  ( x ) = f ( x , q _ 1 , q _ 2 , ... , q _ N )
        \end {equation}
        where the mapping, we notice, is now {\it overcomplete}.

        If we insert this transformation into the Hamiltonian, we naturally
        find, after a straightforward application of the chain rule:

        \begin {eqnarray}
        H  &=& { 1 \over 2 }
        \int dx \, \omega ( x , \phi )
        { \delta \over \delta \phi ( x ) }
        \nonumber \\
        & - & { 1 \over 2 }
        \int dx d y \,
        \Omega ( x , y , \phi )
        {\delta \over \delta \phi ( x )
        \delta \phi ( x ) }
        \end {eqnarray}
        where:

        \begin {eqnarray}
        \omega  ( x , \phi ) & = &
        - \sum _ i \partial _ i ^ 2 f ( x , q _ i )
        \nonumber \\
        \Omega ( x , y , \phi ) & = &
        \sum _ i \partial _i f ( x , q )
        \partial _ i f ( y , q )
        \end {eqnarray}
        Let us write $\pi ( x ) = - i ( \delta / \delta \phi ( x ) ) $.
        Then, symbolically, the free Hamiltonian can be written as:

        \begin {equation}
        H = { 1 \over 2 } \pi \Omega \pi + { i\over 2 }
        \omega \pi
        \end {equation}
        where we have dropped all integral signs for the sake of clarity.

        Unfortunately, this Hamiltonian actually yields incorrect answers
        for even simple systems.
        The reason is that we have not included the Jacobian of the
        transformation, and also because this new Hamiltonian is not hermitian
        on scalar products defined in this new space. The term
        which violates hermiticity is
        $\omega \pi$.

In the old co-ordinates, scalar products are defined via functionals
$\Phi [\phi ]$. In the new co-ordinates, scalar products are
defined via $\Psi [\phi]$, where we have the rescaling $\Psi [ \phi ] =
J ^ { 1/2} \Phi [\phi ]$, where $J[\phi]$ is the Jacobian
of the transformation. This means that expressions
like $\pi \Phi [\phi]$ have to be rescaled via factors
of $ J ^ { - 1/2 }$:

        \begin {eqnarray}
        \pi ( x ) \Phi [\phi]) &=&
\pi J ^ { - 1/2 } \Psi ( [ \phi ] ) \nonumber \\
& = & J ^ { - 1/2 }
        \left (
        \pi - i C \right ) \Psi [ \phi ]
        \nonumber \\
        C & = & - { 1\over 2 } { \delta \ln J [ \phi  ] \over
        \delta \phi ( x ) }
        \end {eqnarray}
        Inserting this new value of $\pi$, shifted by the
quantity $-iC$, we now have, symbolically:

        \begin {eqnarray}
        H &=&
        { i \over 2 }
        \left ( \omega + i ( \pi \Omega ) - 2 C \Omega \right )
        \pi
        \nonumber \\
        & + & { 1 \over 2 }
        \left ( \pi \Omega \pi - C \Omega C \right )
        + { 1 \over 2 } \omega C
        - { i\over 2 } \Omega ( \pi C )
        \end {eqnarray}
        It is now a simple matter to make the Hamiltonian
        hermitian. We will simply set the
        first line of the equation to zero.
This not only restores the hermitian nature
        of the Hamiltonian, as desired, it also determines $J$:

        \begin {equation}
        \omega + { \delta \Omega \over \delta \phi }
        - 2 \Omega C = 0
        \end {equation}
        whose solution is given by:

        \begin {equation}
        C = { 1 \over 2 } \Omega ^ { -1 }
        \left [ \omega + i (\pi \Omega ) \right ]
        \end {equation}

        With this new restriction, our Hamiltonian now
        becomes:

        \begin {eqnarray}
        H &=& { 1 \over 2 } \pi \Omega \pi
        + { 1 \over 8 }
        \left ( \omega +
        { \partial \Omega \over \partial \phi } \right)
        \Omega ^ { -1 }
        \left (\omega + { \partial \Omega \over \partial \phi }\right)
        \nonumber \\
        & + & V ( \phi )
        - { 1 \over 4 }
         { \delta \omega \over \delta \phi }
        - { 1 \over 4 } { \partial ^ 2 \Omega
        \over \partial \phi \partial \phi }
        \end {eqnarray}

        Although this expression appears to be quite formidable,
        it simplifies considerably for concrete systems.
        Now, let us adapt this formalism to matrix models.

        We will make a change of variables from $M(t)$ to
        a new variable $\phi (x , t )$, or its Fourier
        transform $\phi _ k ( t )$, which is given in terms of an
invariant trace:

        \begin {equation}
        \phi_k (t) = \,\, {\rm Tr } \,\, \left( e^{ikM(t)}\right) =
\sum_{i=1}^N \,
        e^{ik\lambda_{i}(t)}\,\,,
        \end {equation}

        \begin {equation}\phi (x,t) = \int {dk\over 2\pi} \, e^{-ikx} \,
        \phi_k (t) =
        \sum_{i=1}^N \, \delta \left( x-\lambda_i (t)\right)\,\,,
        \end {equation}
        We see that $\phi ( x , t )$ is nothing but a
sum over delta functions of the eigenvalues of $M(t)$.
There is, therefore,
an additional constraint $\int dx \, \phi ( x , t ) = N$,
which we can impose by adding a Lagrange multiplier term
$\mu _ F \left( \int dx \, \phi ( x, t ) - N\right )$.
This Lagrange multiplier will correspond to the Fermi level.

        Let us now insert the value of $\phi$ into $\omega$ and
        $\Omega$. A simple calculation yields:

        \begin {eqnarray}
        \omega ( k , \phi ) & = & - { \partial ^ 2 \over \partial M ^ 2 }
        \phi _ k = k ^ 2 \int _ 0 ^ 1  d \alpha
        \phi _ { \alpha k }
        \phi _ { k ( 1 - \alpha )}
        \nonumber \\
        \Omega ( k , k ' , \phi ) & = &
        { \partial \phi _ k \over
        \partial M }
        { \partial \phi _ { k ' }
        \over \partial M }
        = k k ' \phi _ { k + k' }
        \end {eqnarray}

        Written in
        $x$ space, these
        expressions become:
        \begin {eqnarray}
        \omega ( x , \phi )  & = & 2 { \partial \over \partial x }
        \int d y  { \phi ( x ) \phi ( y ) \over x - y }
        \nonumber \\
        \Omega ( x , x ' , \phi ) & = &
        { \partial \over \partial x } { \partial \over \partial y }
        \left [ \delta ( x - x ' ) \phi ( x ) \right ]
        \end {eqnarray}

        Now let us insert the value of $\omega$ and $\Omega$ into the
        Hamiltonian. Our final expression now becomes:

        \begin {equation}
        H =\int dx\,\,\Bigl\{ {1\over 2}\,\Pi_{,x}\,\phi\,\Pi_{,x}+
        {\pi^2\over6} \phi^3 + [ U (x) - \mu _ F ] \phi \Bigr\}\,\,,
        \end {equation}
with the commutation relations:

\begin {equation}
[ \phi ( x ) , \Pi ( y ) ] = \delta ( x - y )
\end {equation}
As before, we have restricted our system to the singlet
sector of the full theory.

Starting with this Hamiltonian, it
is not difficult to make the transition to the Lagrangian formalism
and write down the
        action of the system:

        \begin {eqnarray}
        S & = & \int dt\int dx
        \Bigg [
        { 1 \over 2 }
        \partial _ x ^ { -1 }
        \dot \phi
        { 1\over \phi }
        \partial _ x  ^ { -1 }
        \dot \phi
        \nonumber \\
        & - &
        { 1 \over 6 }
        \pi ^ 2 \phi ^ 3 +
        \left ( \mu _ F - U ( x )  \right )
        \phi ( x , t ) \Bigg  ]
        \end {eqnarray}
        What is rather unusual about this system of
        equations is that the Hamiltonian is
        cubic, but the action is actually non-polynomial.
        This means that there are two distinct ways in which
        to derive the S-matrix of the theory, requiring
        quite different Feynman rules.

        In practice, it will be more convenient, of course,
        to calculate Feynman diagrams with the cubic
        theory, rather than the non-polynomial one.
        To extract out the Feynman rules, we will find  it
        useful to expand around the classical solution of the
        theory:

        \begin {equation}{1\over 2} \left( \pi \phi_0 (x) \right) ^2 + U (x) =
        \mu_{F}\,\,,
        \end {equation}

        \begin {equation}\pi \phi_0 = p_0 (x) = \sqrt{2(\mu_{F} - U(x) )}\,\,.
        \end {equation}

        Then, by power expanding around $\phi _ 0$:
        \begin {equation}
        \phi ( x , t ) =  \phi _ 0 ( x ) +
        { 1 \over {\sqrt \pi } }
        \partial _ x \eta ( x, t )
        \end {equation}
        we find:

        \begin {equation}H = \int dx\,\, \Bigl\{(\pi\phi_0) \Bigl( {1\over 2}\,
        \Pi^2 + {1\over 2}\, \eta_{,x}^2\Bigr) + {\pi^2\over 6}\,
        (\eta_{,x})^3 + {\pi\over 2} \Pi^2 \eta_{,x}\Bigr\}\,\,,
        \end {equation}

In order to eliminate the $\phi _ 0$ in the free part of the Hamiltonian,
we will change variables from $x$ to $\tau$:

\begin {equation}
\tau = \int ^ x { dx \over \pi \phi _ 0 } ;\quad
{ d x ( \tau ) \over d \tau } = p _ 0
\end {equation}
With this new variable, the Hamiltonian becomes,
for the quadratic and cubic pieces:

\begin {equation}
\label{eq:quad}
H _ 2 + H _ 3 = \int d \tau
\left \{
{ 1\over 2 }
\left ( \Pi ^ 2 +
( \partial _ \tau \eta ) ^ 2 \right )
+ { 1 \over 6 p _ 0 ^ 2 }
\left [ ( \partial _ \tau \eta ) ^ 3 +
3 \Pi ^ 2 ( \partial _ \tau \eta ) \right ]
\right\}
\end {equation}

There is also the linear term, given by:

\begin {equation}
\label{eq:lin}
H _ 1 = { 1 \over 12 } \int d \tau
\left [
{ \phi _ 0 ''' \over \phi _ 0 }
-
{ 1 \over 2 }
\left ( { \phi _ 0 ' \over \phi _ 0 } \right ) ^ 2
\right ]
- { 1\over 24 }
\int dx
\left [ { ( \phi _ 0 ' ) ^ 2 \over \phi _ 0 }
- 2 \phi _ 0 ''' \right ]
\end {equation}

{}From this, we can derive Feynman rules and calculate scattering
        amplitudes.
Before we do this, however, we can use the form of the Hamiltonian
found here to reveal some of
the hidden symmetries of this system.

        \section {Discrete States and $w( \infty)$}

        To see that matrix models possess hidden symmetries,
        it is useful to write it in Hamiltonian form:

        \begin {equation}H = {1\over 2}\, \, {\rm Tr } \,\left(P^2-M^2\right)
        \end {equation}
        where $P$ is a matrix which is conjugate to $M$, and we take
$U(M)= - { 1\over 2 } M ^ 2 $.

        Now write down the following
        operators, which correspond to creation--annihilation
        operators:

        \begin {equation}B_n^{\pm} = \, {\rm Tr } \, \left( P \pm M\right)^n\,,
        \quad\quad n = 0,1,2,\ldots
        \end {equation}
        What is remarkable is that these states have the following
        commutation relation with the Hamiltonian:

        \begin {equation}\left[\,H, B_n^{\pm}\,\right]= \mp in
B_n^{\pm}\,,\qquad
        \end {equation}

        This also indicates that these states $B _ n ^ \pm$
correspond to physical states with energy
$\epsilon _ n = \pm i n $.
(We can analytically continue this discrete imaginary momenta
to real values, so that this operator corresponds to a
physical state.)

        We can also generalize this operator
and write down
another series of operators:

        \begin {equation}B_{n,\bar n}=\, {\rm Tr } \,\Bigl((P+M)^n
(P-M)^{\bar{n}}\Bigr)\,\,,
        \end {equation}
        which has the commutation relations:

        \begin {equation}\left[ H, B_{jm} \right] = -2i m B_{jm}\,\,.
        \end {equation}
        where:
        \begin {equation}m ={n-\bar{n}\over 2}\,,\qquad\quad
        j ={n + \bar{n}\over 2} \,\,.
        \end {equation}
These states have energy given by $ - 2 i m $.

        We would now like to
reveal the physical meaning of these operators.
The easiest way to do this is to
carry over these expressions to
the
collective field theory.
        In this way, we will be able to establish the field-theoretic
        meaning of these operators.

        To do this, the simplest method is to construct a
        \lq\lq dictionary" which will give us a quick,
        intuitive way in which to make the transition between
        matrix model quantities and collective field theory.
        To construct this dictionary, we will introduce
        the operator:

        \begin {equation}
        \alpha _ \pm ( x , t ) = \Pi _ { ,x }
        \pm \pi \phi ( x ,t )
        \end {equation}
        which has the commutation relations:

        \begin {equation}\left[
 \alpha_{\pm} (x) , \alpha_{\pm} (y) \right] = \pm 2\pi
        \partial_x \delta (x-y)
        \end {equation}

        Written in this form, the Hamiltonian can be written as:

\begin {equation}H  = {1\over 2} \int \, {dx\over 2\pi}\,\Bigl\{
        {1\over 3} \left( \alpha_+^3 - \alpha_-^3\right) - \left( x^2 - \mu
        \right) \left( \alpha_+ - \alpha_- \right) \Bigr\}\,\,.
        \end {equation}

        To make the transition between matrix models  and
        collective field theory, the matrix $M$ is eventually replaced
        by its eigenvalues $\lambda$, which eventually
        become the variable $x$. Also, the conjugate matrix
        $P$ is replaced by its eigenvalues $p$, which becomes the operator
        $\alpha ( x, t )$.

        \begin {eqnarray}
        M & \leftrightarrow & \lambda \leftrightarrow x\,\,,
        \nonumber \\
        P &\leftrightarrow & p \leftrightarrow \alpha (x,t)\,\,.
        \end {eqnarray}

        Furthermore, one can show that the trace operation
        is replaced by:

        \begin {equation}\, {\rm Tr } \, \left\{ \,\,\right\}\quad \rightarrow
\quad\int{dx\over 2\pi}
        \int_{\alpha_{-}(x,t)}^{\alpha_{+}(x,t)} d\alpha \,\, \left\{ \,\,
        \right\}\,\,,
        \end {equation}
        where $\alpha_{\pm} (x,t)$ are the chiral components of the scalar
        field density. For example, one can show that the matrix
        model Hamiltonian
        is easily transformed into the collective field theory Hamiltonian
        by these substitutions:

        \begin {equation}{1\over 2}\,\, {\rm Tr } \, \left( P^2 - M^2
\right)\rightarrow{1\over 2}
        (p^2 - x^2) \rightarrow \int {dx\over 2\pi} \int d\alpha \,
        {1\over 2}(\alpha^2- x^2) ={1\over 2} \int {dx\over 2\pi}
        \Bigl[ {\alpha\over 3}^3 - x^2 \alpha \Bigr]_-^+ \,\,.
        \end {equation}

        With this dictionary, we can now find the meaning of the
        operator $B _n$ and $B _ { j,m}$.
        The $B _ n$ operators simply
correspond to the tachyons of the theory.

        Also, consider the operator:
        \begin {equation}H_n = {1\over 2\pi} \int dx \int_{\alpha_-
(x,t)}^{\alpha_+ (x,t)}
        d\alpha \,\, \left( \alpha^2 - x^2 \right)^n\,\,,
        \end {equation}
        which possess the commutation relations:
        \begin {equation}\left[ H_n , H_m \right] = 0\,\,,
        \end {equation}
        which are conserved:

        \begin {equation}{d\over dt}\, H_n = \int dx\,\partial_x \left(
\alpha^2 - x^2 \right)
        \left( \alpha^2 - x^2 \right)^n = 0\,\,.
        \end {equation}
        This shows that the Hamiltonian is nothing but a member of
        a spectrum-generating algebra.

These elements, in turn, can be generalized to the larger set:

        \begin {equation}O_{JM} = \int {dx\over 2\pi}
\int_{\alpha_{-}}^{\alpha_{+}} d\alpha
        \,\, (\alpha + x )^{J+M+1} (\alpha - x )^{J-M+1}\,\,,
        \end {equation}
        which obeys the
        $w(\infty)$ commutation relations:$^{15}$

        \begin {equation}\left[ O_{J_{1}M_{1}} , O_{J_{2} M_{2}} \right] = 4i
        \Bigl((J_2+1)M_1-(J_1+1)M_2\Bigr)\,O_{J_{1}+J_{2},M_{1}+M_{2}}\,\,.
        \end {equation}

The states $ B _ { jm }$ therefore correspond, in
collective field theory language, to the discrete
states.
They are physical states of the theory, but they differ
physically from the tachyon states $B _ n$ because they are defined
only at discrete momenta.

        In summary, we found
that the states $B _ n$ and $B _ {jm }$
correspond, respectively, to the tachyon
and the discrete states once we translate
our operators into collective field theory
language via this \lq\lq dictionary."
        We also found that the Hamiltonian is nothing but a single
        member of a spectrum-generating algebra, given by
        $w(\infty)$.
        The fact that we have an infinite number of conserved currents
        is yet another way to see that the model is soluble.

        \section {Equivalence with Free Fermion Model}

At this point, it may seem strange that we have two entirely
different formalisms in which to describe $D=2$ string theory,
the free fermion theory and the collective field theory.
Actually, they are equivalent.
To demonstrate this fact, we will find it useful to bosonize
the free fermion fields in eq. (\ref{eq:fermion})
in terms of bosonic variables
$P$ and $X$ as follows:$^6$

        \begin {eqnarray}
        \Psi_L &=&{1\over\sqrt{2\pi}}
        :\exp\big[i\sqrt\pi\int^{\tau} (P - X')d\tau'\big] :\ ,
        \nonumber\\
        \Psi_R &=&{1\over\sqrt{2\pi}}
        :\exp\big[i\sqrt\pi\int^{\tau} (P + X')d\tau'\big] :\ ,
        \end {eqnarray}
        where $X$ is a massless two-dimensional periodic scalar field,
        and $P$ is its canonically conjugate
        momentum.

It is now a simple matter to insert these bosonized expressions into
the action of the
free fermion theory. We find, for example, expressions like:

        \begin {eqnarray}
        :\Psi_L^{\dagger}\partial_{\tau}\Psi_L- \Psi_R^{\dagger}\partial_{\tau}
\Psi_R:
        &=&
        {i\over 2} : P^2 + (X')^2 :
         \nonumber\\
        :\Psi_L^{\dagger}\Psi_L+ \Psi_R^{\dagger} \Psi_R: &=&
- {X'\over\sqrt\pi} \nonumber\\
        :\partial_{\tau}\Psi_L^{\dagger}\partial_{\tau}\Psi_L+
        \partial_{\tau}\Psi_R^{\dagger}\partial_{\tau} \Psi_R:
        &=&
         -\sqrt\pi :P X'P + {1\over 3} (X')^3 +
        {1\over 6\pi} X''':  \,\,\,\,.
        \end {eqnarray}

        It is now straightforward to substitute these
expressions directly into the Hamiltonian:

        \begin {eqnarray}
        :  H : & = &{1\over 2}\int_0^{T/2} d \tau :\bigg[ P^2 + ( X')^2
        - {\sqrt\pi \over \beta v^2} \bigl(PX'P + {1\over 3} (X')^3
        + {1\over 6\pi} X'''\bigr)\nonumber\\
        & - & {1\over  2\beta \sqrt\pi} X'\left({v''\over v^3}-{5 (v')^2 \over
2 v^4}
        \right)\bigg]:
        \,.
        \end {eqnarray}
where $ v = \sqrt { 2 ( \mu _ F - U ) } $.
        If we integrate by parts and discard the boundary terms,
        this reduces to:

        \begin {equation}
        :  H : = {1\over 2}\int_0^{T/2} d \tau :\bigg[ P^2 + ( X')^2
        - {\sqrt\pi \over \beta v^2} (PX'P + {1\over 3} (X')^3 )
        -{1\over  2\beta \sqrt\pi} X'\left({v''\over 3v^3}-{(v')^2 \over 2 v^4}
        \right)\bigg]:
        \,.
        \end {equation}

 In this way, we can now directly compare the free fermion action
with the action of collective field theory in
eqs. (\ref{eq:quad}) and (\ref{eq:lin}),
and we find that
they are identical.
Physically, the massless field $X$ describes
        small fluctuations of the Fermi surface.
        In the Hamiltonian, this reduces to (for $\mu>0$
and $v = {\sqrt { 2 \mu } } \,{\rm sinh } \, (\tau) )$:

        \begin {equation}
        : H : = {1\over 2}\int_0^\infty d \tau :\bigg[ P^2 + ( X')^2
        - {\sqrt\pi \over 2\beta \mu\sinh^2 \tau} \bigl(PX'P +
         {1\over 3} (X')^3 \bigr )
        -{1-{3\over 2}\coth^2\tau \over  12\beta \mu\sqrt\pi \sinh^2\tau}
        X' \bigg]:
        \,.
        \end {equation}

There is one subtle point. Like the free fermion action, this Hamiltonian
is actually divergent at $\tau=0$. This divergence can be
regularized
        with zeta-function techniques.
        Alternatively, we may take this with $\mu<0$. Then

        \begin {eqnarray}
        : H : & = & {1\over 2}\int_0^\infty d \tau :\bigg[ P^2 + ( X')^2
        - {\sqrt\pi \over 2\beta |\mu| \cosh^2 \tau} \bigl( PX'P + {1\over 3}
(X')^3 \bigr )
        \nonumber \\
        & -& {1-{3\over 2}\tanh^2\tau \over  12\beta |\mu|\sqrt\pi \cosh^2\tau}
        X' \bigg]:
        \ ,
        \end {eqnarray}

If we compare this result with the previous one derived from
collective field theory, then we see that they are the same.
Now that we have established the equivalence of the two theories,
we can now
use this expression to derive the Feynman rules and then
the amplitudes.

\section {S-matrices}

We decompose the $X$ and $P$ fields into oscillators as
follows:

        \begin {eqnarray}
        X(t, \tau) & =& \int_{-\infty}^\infty
         {dk\over\sqrt {4\pi |k|}}
        \left (a(k) e^{i(k\tau-|k| t)}+a^{\dag} (k)
        e^{-i(k\tau-|k| t)}\right ) ,
        \nonumber\\
        P(t, \tau)& =& -i\int_{-\infty}^\infty {dk\over\sqrt {4\pi |k|}}|k|
        \left (a(k) e^{i(k\tau-|k| t)}-a^{\dag} (k)
        e^{-i(k\tau-|k| t)}\right )\ ,
        \end {eqnarray}
        such that $[a(k), a^{\dagger}(k')]=\delta (k-k')$.
The Hamiltonian now decomposes into three pieces:

        \begin {eqnarray}
        H_2 & =& \int_0^\infty dk k \, a^\dagger (k) a(k)\ ,
        \nonumber\\
        H_3 & =&
        {i\over 24\pi\beta |\mu|} \int_0^\infty dk_1 dk_2 dk_3\sqrt{k_1 k_2
k_3}
        \bigl [f(k_1+k_2+k_3) a(k_1) a(k_2) a(k_3)
        \nonumber\\
        &- & 3f(k_1+k_2-k_3) :a(k_1) a(k_2) a^\dagger (k_3):\bigr ]
        +h. c. \ ,
        \nonumber\\
        H_1 & =& -{i\over 48\pi\beta|\mu|}
        \int_0^\infty dk \sqrt k g(k) \bigl(a(k)-a^{\dagger}(k)\bigr)\ ,
        \end {eqnarray}
where $H _ 2$ is the free Hamiltonian, $H _ 3$ is the cubic interaction,
and $H _1$ is the tadpole term, and:

        \begin {eqnarray}
        f(k) &=& \int_{-\infty}^\infty d\tau {1\over \cosh^2\tau} e^{ik\tau}
        ={\pi k\over \sinh (\pi k/2)}\ ,
        \nonumber\\
        g(k) & =& \int_{-\infty}^\infty
        { 1-{3\over 2}\tanh^2\tau
        \over  \cosh^2\tau} e^{ik\tau}=
        {\pi (k^3+2k)\over 4\sinh (\pi k/2)}
        \ .\nonumber\\
        \end {eqnarray}
One novel feature is the appearance of
the functions $f$ and $g$, which appear in the three-tachyon
and one-tachyon Hamiltonian.
When calculating scattering amplitudes, they give
rise to an infinite tower of
poles at discrete momenta, i.e. they will represent
the effect of the discrete states on the scattering amplitudes.

        Now we calculate the $S$-matrix:
        \begin {equation}S=1-2\pi i\delta (E_i-E_f) T\ . \end {equation}

Each right-moving massless particle has energy equal to momentum,
        $E=k$.
Let us describe the case of non-forward
scattering of particles of momenta $k_1$ and $k_2$ into particles
        of momenta $k_3$ and $k_4$. Second-order
        perturbation theory gives us:

        \begin {equation} T(k_1, k_2; k_3, k_4)=
        \sqrt{E_1 E_2 E_3 E_4}
         \sum_i {<k_3 k_4|H_{int}|i><i| H_{int}|k_1 k_2>\over E_1+E_2-E_i}
        \ ,\end {equation}
        where we sum over all intermediate states $i$.
        The $s$-channel contribution is then given by:

        \begin {equation}
        S^{(s)}  (k_1, k_2; k_3, k_4)=-i
        {g^2_{st}\over 8\pi} \prod_{j=1}^4 E_j
        \int_0^\infty dk \left (
        {k f^2 (k_1+k_2-k)\over k_1+k_2-k+i \epsilon}
        -{k f^2 (k_1+k_2+k) \over k_1 + k_2 + k - i\epsilon} \right )\ ,
        \end {equation}
The first term is due to one-particle intermediate exchange. The
second one is due to five-particle intermediate states (found
by slicing the four-particle scattering amplitude in a particular
fashion).
Likewise, the $t$- and $u$-channel contributions can also be
calculated. By adding all three channels together,
we find:$^{16}$

        \begin {equation} S(k_1, k_2; k_3, k_4)=-i{\pi g_{st}^2\over 8}
        \delta (E_1+E_2-E_3-E_4)\prod_{j=1}^4 E_j\bigl (F(p_s)+F(p_t)
        +F(p_u)\bigr )
        \ ,\end {equation}
        where $p_s=k_1+k_2$, $p_t=|k_1-k_3|$, $p_u=|k_1-k_4|$, and

        \begin {equation}
        F(p)=\int_{-\infty}^\infty dk \left
        [{(p-k)^2\over \sinh^2 (\pi (p-k)/2)}\right ]
        {k \over p-k+i\epsilon ~ {\rm sgn} (k)}
        \end {equation}

        Using:
        \begin {equation}{1\over x-i\epsilon }={\cal P}{1\over x}+\pi i\delta
(x)
        \end {equation}
        we find
        $F(p)=-i (4p / \pi ) - ( 8 / 3 \pi) $.
        The total amplitude thus becomes:

        \begin {equation} S(E_{1,2}; E_{3,4})=-{g_{st}^2\over 2}
        \delta (E_1+E_2-E_3-E_4)\prod_{j=1}^4 E_j\bigl (
        E_1+E_2+|E_1-E_3|+|E_1-E_4|-2i\bigr )
        \ . \end {equation}

        Upon the Euclidean continuation $E_j\to i|q_j|$,
        and inclusion of the external leg factors,
        this agrees with the calculation found in matrix models.

This amplitude is rather revealing for several reasons.
The amplitude seems to consist of two parts, the momentum-dependent
parts, given by $|k _ 1 + k _ 2 |$, etc.
and a contact term, given by a constant.
The momentum-dependent terms, in turn, correspond
to the exchange of the tachyon.
The contact term, on the other hand, emerges because we must
sum over an infinite number of poles in the
$F (p)$ integration. These poles, we see, correspond to the
discrete states which are defined at discrete momenta.

Thus, it is possible to define a second set of Feynman rules,
seemingly different from the previous set,
in which we explicitly reveal the presence of
the tachyon exchange and the presence of an infinite tower
of discrete states.
In contrast to the previous set of Feynman rules, which
are based on cubic interactions, the second set of Feynman
rules are actually non-polynomial, since
the contact interactions can be defined over $N$ particle
states.

After a simple rescaling, we can define this new set of
Feynman rules as the tachyon propagator and the
various contact vertex functions $V _ N$
arising from the discrete states:

\begin {eqnarray}
{\rm tachyon \, propagator} & \rightarrow & i | k |
\nonumber \\
V _ 3 & \rightarrow & { 1 \over \mu}
\nonumber \\
V _ 4 & \rightarrow & { 4 \over \mu ^ 2 }
\nonumber\\
V _ 5 & \rightarrow & { 1 \over \mu ^ 3 }
\left( 32 + \sum _ { i = 1 } ^ 5 k _ i ^ 2 \right)
\end {eqnarray}

The existence of two sets of Feynman rules, one based on
cubic vertices, and the other being non-polynomial,
is apparently a universal feature of string theory.
On one level, this is due to the fact that the
collective field theory has an non-polynomial
Lagrangian formulation as well as a
cubic Hamiltonian formulation.
Similarly, ordinary string field theory in 26 dimensions
also has
the cubic light-cone vertex function as well as the
covariant non-polynomial form, both of which yield
the same on-shell S-matrix.

As an added check, one can show that these Feynman rules,
both cubic and non-polynomial, can be
shown to agree with the usual value of the $N$ point
amplitudes, which equals,
after a rescaling in a special kinematical region:$^{17}$

\begin {equation}
\label{eq:mu}
T _ N \sim
\left( { d \over d \mu} \right ) ^ { N-3}
\mu ^ { s + N - 3 }
\end {equation}
where:

\begin {equation}
s = { 1\over \sqrt 2 }
\sum _ { i = 1 } ^ N | q _ i |
- ( N - 2 ) ; \quad
\sum  _ { i = 1 } ^ N q _ i = 0
\end {equation}
where $ q _ i = - k _ i / 2 $.

\section {Temporal Gauge String Field Theory}

In the usual 26 dimensional critical string
field theory, there are both cubic and non-polynomial
actions. The
cubic theory is associated with the light-cone gauge,
where the
cubic interaction corresponds to the splitting
or fissioning of
a closed string into two smaller pieces, conserving
string length in the process, i.e. $l _1 + l _ 2 = l _ 3$.
The 26 dimensional non-polynomial theory, by contrast,
preserves the string length of each string, and hence
the non-polynomial interactions do not conserve string length.

Similarly, collective field theory in two-dimensions
has both a
cubic and a non-polynomial
form. One puzzling aspect of the collective field theory
approach, however, is that the cubic
version does not have any remnant of the string length.
In fact, no where in the collective field theory does the
string length enter.

Some light on this peculiar situation is shed by the
temporal gauge string field theory, where the
world-sheet graviton is quantized in the temporal gauge.
It was found by Ishibashi and Kawai$^8$ that the
quantization of a pure $c=0$ two-dimensional gravity theory
(in zero space-time dimensions) yielded a purely cubic
action, where the string length was preserved.
They showed that the correlation functions of the
temporal gauge gravity theory were identical to the
one-matrix model theory by demonstrating that the
$\tau$ functions$^{18}$ of the theory
obeyed the usual Virasoro conditions.

In two-dimensional gravity in this gauge, the physical states
are labeled by the invariant string length, which is a generally
covariant
quantity, and hence interactions
preserve string length.
They then showed that this formalism generalized to the
multi-matrix approach, showing that the
$ c \leq 1$ case also preserved string length in its cubic
interactions.
In the limit as $c \rightarrow 1$, one presumably approaches
the two-dimensional string theory.

Perhaps the simplest way in which to derive this new
string field theory is to apply the apparatus of
collective field theory to the
case of the one-matrix model, and then let the number of
matrices go to infinity.

The multi-matrix approach$^{19}$ begins with a finite sequence of
hermitian matrices $M _ i$, with an action:

\begin {equation}
S = -c \sum_i \, Tr (M_i M_{i+1} ) + \sum_i \, V_i (M_i )
\end {equation}

Correlation functions can be computed from the functional integral:

\begin {equation}Z = \int dM_1 \cdots dM_k e^{-S}
\end {equation}
with hermitian $N\times N$ matrices $M_i$, $i=1, \cdots, k$.

It can be shown that this generates a sequence of
matrix models such that $c$
is given by:
\begin {equation}
c  = 1- {6 \over (k+1)(k+2)} , k = 1 \cdots n
\end {equation}
For $k=1$, we have the one-matrix model with $c = 0$.
For $k \rightarrow \infty$, we presumably have the
$c = 1$ model.
In the double scaling limit,
these theories
correspond to two-dimensional gravity coupled
to conformal matter with various values of $c$.

The $U(N)$ invariant observables can be given by the
traces

\begin {equation}\phi_C = \, {\rm Tr } \,  (M_1^{n_{1}} M_2^{n_{2}} \cdots )
\end {equation}
where the loop space index $C$
denotes the sequence of matrices
$C = \{ n_1 , n_2 , \cdots  \}.$

There is one important complication, however.
There are no dimensions of space-time present.
This means that the Hamiltonian makes no sense, because there is no
time parameter.
To remedy this situation, we will use the stochastic quantization
approach$^{20}$ and introduce a fictitious time variable $t$ into the
theory and take the limit as $t \rightarrow \infty$.

In the stochastic quantization approach, we begin with
an action $S$ defined with a field $\varphi (x )$ and
introduce a fictitious time variable $t$. The
field $\varphi ( x , t )$ obeys the time-dependent
Langevin equation:

\begin {equation}{\partial\over\partial t} \varphi (x,t) = - {\partial
S\over\partial\varphi}
+ \eta\end {equation}
where $\eta$ is the random variable.

The correlation functions are then
obtained in the limit $t \rightarrow \infty$
\begin {equation}\langle F(\varphi )\rangle = \lim_{ t\rightarrow\infty}
\int \left[ d\varphi (x) \right] F(\varphi ) P_t\end {equation}
where the time evolution (of the probability distribution
$P _ t$) is given
by
 \begin {equation}{\partial\over \partial t} P_t = - H_{FP} P_t\end {equation}
with the Fokker-Planck Hamiltonian
\begin {equation}H_{FP} = -{1\over 2} \int \left( {d\over d\varphi (x)} -
{\delta S\over
\delta \varphi (x)} \right) {d\over
d\varphi(x)}\end {equation}

In the limit $t \rightarrow \infty$, one can show that these
stochastic correlation functions defined over
a fictitious time
approach the correct correlation functions.

Our task is now to re-write these equations in terms of the matrix
model approach. To begin, let us start with the
$c = 0$ one-matrix model.
As in the $c=1$ case, we will
make the usual change of variables from $M$ to $\phi _ n$,
such that $\omega$ and $\Omega$ become, by the chain rule:

\begin {eqnarray}
\phi_n & = & \, {\rm Tr } \,  (M^n )\nonumber \\
 \Omega (n,n^{\prime})& = & nn^{\prime} \phi_{n+n'-2}\nonumber \\
 \omega (n) & = & -n \sum_{n'=0}^{n-2} \, \phi_{n'} \phi_{n-n'-2}
\end {eqnarray}

The stochastic Hamiltonian and action are given by:$^{21}$

\begin {equation}S = Tr \left( {1\over 2} \mu M^2 - {1\over 3} M^3 + \cdots
\right)
\end {equation}
and
\begin {equation}H = - Tr \left( {\partial\over\partial M} - {\partial S\over
 \partial M}\right) \, {\partial\over\partial M}\end {equation}

Making the standard change of variables, the
Hamiltonian becomes
\begin {equation}
- \sum_n \left( \sum_{m=0}^{\infty} \phi_{n+m-2} m {\partial
\over\partial \phi_m} + \sum_{r=0}^{n-2} \phi_r \phi_{n-r} -
 \Omega \left( S, \phi_n \right) \right) n {\partial\over \partial
\phi_n}\end {equation}
with
\begin {equation}\Omega (S,\phi_n ) = n \left( \mu \phi_{n+1} - \phi_{n+2} +
\cdots
\right)\end {equation}

(This Hamiltonian has an elegant structure. If we drop the
$S$ dependent term, then it can be
written as
\begin {equation}H_3 = - \sum_n \, O_n \, n\Pi_n \end {equation}
with
\begin {equation}O_{n+2} = \sum_{n=0}^{\infty} \, \phi_{n+m} m \Pi_m + \sum_r
\phi_r
\phi_{m-r}
\end {equation}
where we have a Virasoro algebra generated by $O_{n+2} = L_n$:
\begin {equation}\left[ L_n , L_m \right] = \left( n-m\right) L_{n+m}\end
{equation}

The Hamiltonian is therefore of the form
$H = - O _ n n ( d / d \phi _ n )$.)

Let us now take the continuum limit. We will take the
$z$ representation, given by:

\begin {equation}\phi (z) = {\rm Tr} \, {1\over z-M} = \int_0^{\infty}
dLe^{-Lz} \phi_L =
\int_0^{\infty} dLe^{-Lz} Tr (e^{LM})\end {equation}
with:

\begin {eqnarray}
\phi(z)& = & \sum_{n\geq 0} \, z^{-n-1} \phi_n\nonumber \\
\partial \Pi (z) & = & \sum_{n\geq 0} \, z^{n-1} n \Pi_n
\end {eqnarray}
with:
\begin {equation}O(z) =\sum_n \, z^{-n-2} O_n\end {equation}
This gives:
\begin {equation}O(z) = \int dz^{\prime} {\phi(z') - \phi (z)\over z - z'}
 \, \partial_{z'} \Pi + \phi^2 (z) \equiv (\phi \partial_{z}\Pi )(z) + \phi^2
(z)\end {equation}
(The bracket notation means that there are only $z^{-n}$
 components.)
The Hamiltonian is now
\begin {equation}H = - \int dz \left[ \phi (z) \partial_z \Pi (z) + \phi^2
(z) + \left( z^2 - \mu z\right) \phi + \left( \mu - z
\right) + c_0 \right] \, \partial_z \Pi (z) .\end {equation}
The scaling limit is defined by $\mu =  \mu_c + a^2 \Lambda
$ and $z  =  z_a + a \zeta  $.
We will also make a shift:

\begin {equation}\phi (z) = {1\over 2} (z\mu - z^2 ) + a^{3/2} \Phi
(\zeta)\end {equation}
where $a$ represents a basic dimensional length.
After a rescaling and a shift in the field, the Hamiltonian
now becomes:

\begin {equation}H = - a^{1/2} \int d\zeta \left[ {1\over N^2 a^5} \, \Phi
 \partial_{\zeta} \Pi + \Phi (\zeta )^2 - \Phi_0^2 \right]
 \partial \Pi (\zeta )\end {equation}

So far, the
Hamiltonian resembles the one found for $c=1$ matrix models.
But the next crucial step is to
convert variables from $\zeta$ to
$l$ (which will correspond to string length):

\begin {equation}\Phi (\zeta ) = \int_0^{\infty} dl \,\,e^{-l\zeta} \Phi
(l)\end {equation}

Inserting this new field definition in terms of string length,
we now find:$^{21}$

\begin {eqnarray}
H &= & - \int_0^{\infty} dl_1 \int_0^{\infty} dl_2
\Big [ g \Phi \left( l_1 + l_2 \right) l_1 \Pi (l_1 ) l_2 \Pi (l_2 )
\nonumber\\
& + &
 \left( l_1 +l_2 \right) \phi (l_1) \Phi (l_2) \Pi (l_1 + l_2 )
\Big ]
\nonumber \\
&-&  \int_0^{\infty} dl\,\, \rho_0 (l) \Pi (l)
\end {eqnarray}

If we examine this Hamiltonian, we find that the string length
$l$ is preserved by the cubic interaction, i.e. closed
strings simply break or fission.
This is the action found by Ishibashi and Kawai.
(In some sense, the fact that the action is now written
in terms of fields which conserve string length is not
surprising. If we have the product of three functions defined
at the same point $\zeta$ and then re-express each function
in terms of its Fourier transform with variable $l$,
then the new expression will
conserve $l$ among the three transformed functions.)

The next step is to generalize these arguments to the
$c \leq 1$ case.
In principle, all steps can be carried out as before, although
there is some difficulty reproducing the constraints on the
$\tau$-functions, which should obey the $w (n)$ algebra constraints.

For the general case, we add an index $i$ to the field $\Phi$,
which corresponds to the group theoretical height of the representation,
where the height variable takes on its value on the nodes of an
$ADE$ or $\hat A\hat D\hat E$ Dynkin diagram.
The world-sheet is divided into different neighborhoods,
where the height variable takes on the same value in the
same neighborhood.

The Hamiltonian is then given by:$^8$

\begin {eqnarray}
H & = & \sum _ i \int _ 0 ^ \infty d l _ 1
\int _ 0 ^ \infty d l _ 2
\, \,
\Phi _ i ^ \dagger ( l _ 1 )
\Phi _ i ^ \dagger ( l _ 2 )
\Phi _ i ( l _ 1 + l _ 2 ) ( l _ 1 + l _ 2 )
\nonumber \\
& + &
\sum _ { i,j}
C _ { ij }
\int _ 0 ^ \infty
dl _ 1
\int _ 0 ^ \infty
d l _ 2
\, \,
\Phi _ i ^ \dagger
( l _ 1 + l _ 2 )
\Phi _ j ^ \dagger ( l _ 2 )
\Phi _ i ( l _ 1 ) l _ 1
\nonumber \\
& + &
g \sum _ i \int _ 0 ^ \infty
d l _ 1 \int _ 0 ^ \infty
d l _ 2
\, \,
\Phi _ i ^ \dagger ( l _ 1 + l _ 2 )
\Phi _ i ( l _ 1 ) \Phi _ i ( l _ 2 ) l _ 1 l _ 2
\end {eqnarray}
where $C _ { ij }$ is the connectivity matrix
($C _ { ij } = 1 $ when $i$ and $j$ are linked
on the Dynkin diagram; otherwise, it vanishes).

In principle, this new Hamiltonian in the $c \rightarrow
\infty$ limit should approach the collective string field
theory Hamiltonian when expressed in terms of the variable $l$,
although this final step has not yet been completed.

        \section {Liouville Theory}

        Several features of collective field theory
        are still rather mysterious. The algebra
        $w(\infty)$ does not emerge as part of
        a gauge symmetry, but rather as a spectrum generating
        algebra.
        Also, the meaning of the discrete states is rather obscure.
        The reason why collective field theory does not resemble a
        standard gauge theory is because all the string degrees of
        freedom have been eliminated, leaving only the tachyon.
        It is thus instructive to re-write all our results in
        a standard Liouville framework.

        Our starting point is the Liouville action
in eq. (\ref{eq:louie}),
with
zero cosmological constant $\mu = 0$, which gives us the
        energy-momentum  tensor:
        \begin {eqnarray}
        T^{(X,\phi)}_{zz}\ & =& \  - { 1 \over 2 }
        (\partial _z X^i)^2 - { 1 \over 2 }  (\partial _z\phi)^2
        + { 1 \over 2 }  Q \partial ^2_z \phi
        \nonumber \\
        T^{(b,c)}_{zz}\ & = & \ -2 b_{zz} \partial _z c^z + c^z \partial _z
b_{zz} \ .
        \nonumber \\
        \end {eqnarray}

        The Fourier components are given  by:
        $L_n = { 1 \over 2 \pi i }
        \oint dz\, z^{n+1} T^{(X,\phi,b,c)}_{zz}$, which form the Virasoro
algebra

        \begin {equation}
        [L_n,L_m]\ =\ (n-m)L_{n+m} + {c+1 +3Q^2 - 26 \over 12} \, n(n^2-1)\,
        \delta_{n+m,0}\ .
        \end {equation}

        To cancel the ghost contribution  to the conformal anomaly one must
        set $c+1 + 3 Q^2 - 26 =0$. Therefore, we have:

        \begin {equation}
        Q\ =\ \sqrt{{25-c\over 3}}\ .
        \end {equation}
For $c =1$, we have $Q = 2 \sqrt 2$.

Let us analyze
these states from the perspective of BRST string field theory.
We start with the BRST operator:

        \begin {equation}
        Q_{BRST}\ =\ \frac1{2\pi i} \oint dz \, c(z)\left( T^{(X,\phi)}(z) +
        {1 \over 2 }  T^{(b,c)}(z) \right)\ .
        \end {equation}

Then we introduce the
field functional $\Phi ( X , \phi , b  , c )$,
where $X$ is the string variable, $\phi$ is the Liouville
field coming from the metric, and $b$ and $c$ are the
Faddeev-Popov ghosts.

Then the free field action is given by:

\begin {equation}
\langle \Phi | Q ( b _ 0 - \bar b _ 0 ) | \Phi \rangle
\end {equation}
which is invariant under: $\delta | \Phi \rangle =
Q | \Lambda \rangle$. The equations of motion are then given by:
$Q ( b _ 0 - \bar b _ 0 ) | \Phi \rangle = 0$.

The physical, on-shell spectrum is found by solving the usual
constraints and equations of motion.
We first note that we can use the gauge degree of freedom to
eliminate all the $b$ and $c$ modes in the field functional
$| \Phi \rangle$.
Then the ghost-free field functional obeys the usual Virasoro conditions:

        \begin {eqnarray}
        L_n | \psi \rangle \ & =& \ \overline{L}_n
        | \psi \rangle \ =\ 0 \qquad\qquad {\rm
        for}\ n>0\
        \nonumber \\
        L_0 | \psi \rangle \ & =& \ \overline{L}_0
        | \psi \rangle \ =\ 1\,\cdot\,
        | \psi \rangle \ . \nonumber \\
        \end {eqnarray}

        Now let us find the lowest physical state.
The ground state of the theory, given by a tachyon,
        is:

        \begin {equation}
        | p, \epsilon \rangle =\ e^{ipX+\epsilon\phi}(0)
        | 0 \rangle \ .
        \end {equation}

The tachyon state satisfies $L_n | p, \epsilon \rangle
        = \overline{L}_n | p, \epsilon \rangle =
        0$, $n>0$,  and

        \begin {equation}
        L_0 | p, \epsilon \rangle  = \overline{L}_0
        | p, \epsilon \rangle \ = \ \left[
        {1 \over 2 }  p^2 -{1 \over 2 }  \epsilon(\epsilon+2\sqrt2)\right]
        | p, \epsilon \rangle \ .
        \end {equation}

        In order to maintain the on-shell condition,
this means that $p^2 -
        \epsilon(\epsilon+2\sqrt2)=2$ or, defining $E=\epsilon +\sqrt2$,
        \begin {equation}
        p^2 -E^2\ =\ 0\ .
        \end {equation}
        meaning that the tachyon is massless.

If we solve for $\epsilon$, we find the rather unusual
solution:

\begin {equation}
\epsilon = - { \sqrt 2 } \pm p = - {\sqrt 2 } + \chi p
\end {equation}
For the case of zero cosmological constant, $\mu = 0$,
this means that
on-shell tachyons occur in two chiralities, $ \chi = \pm 1 $.
When we calculate scattering amplitudes in the next section,
we must keep
track of these chiralities.
(For the case of non-vanishing cosmological constant,
the situation is much more involved. Although the
Virasoro generators are quite non-linear, one can still
show that the Virasoro algebra is preserved$^{22}$ and that
the tachyon state is a solution of the
Virasoro constraints, but only for one value of the
chirality.)

        \section {Tachyon Correlation Functions}

Next, we would like to calculate the scattering amplitudes
for tachyons.
The calculation using string field theory is rather involved and
will be presented later.
We will thus first evaluate the tachyon scattering
amplitudes using conformal field theory.
The tachyon vertex is given by:

\begin {equation}
T _ \chi ( p ) = e ^ { i pX + ( - {\sqrt 2 } + \chi p ) \phi }
\end {equation}
so the correlation function becomes:

        \begin {equation}
        \Big \langle {\prod_{i=1}^N T_{\chi_i}(p_i) } \Big \rangle \ .
        \end {equation}

When performing the functional integral in this correlation
function, there is an important complication.
Since the theory is translationally
        invariant in $X$, the correlator is non-vanishing only if
        $\sum_i p_i =0$.
However, the correlation function is not translation invariant
in $\phi$, which means that $\epsilon$ is not
        conserved, even for vanishing cosmological constant.
In particular, this means that we have difficulty
performing the functional integration over the zero mode $\phi _ 0$,
where $\phi = \phi _ 0 + \tilde \phi$.

In order to integrate over $\phi _ 0$,
we must impose yet another condition on the momenta.
We will demand that all terms proportional to $\phi _ 0$ in the
integrand vanish.
If we set $\mu = 0$, then the $\phi_0$
        dependence of the integrand is given by
the vertices $e ^ { \epsilon _ i \phi _ 0}$ as well
as the curvature term in the action in eq. (\ref{eq:louie}):

        \begin {equation}
        {\rm exp}\left[\sum_{i=1}^N \epsilon_i \phi_0 + \frac1{8\pi} \int d^2
        \sigma \sqrt{\hat{g}} Q \hat{R} \phi_0 \right]\ =\ {\rm exp} \left[
        \left( \sum_{i=1}^N \epsilon_i + 2\sqrt2 \right) \phi_0 \right]\ .
        \end {equation}

        In order to make this vanish, we will demand that:
        \begin {equation}
        \sum_{i=1}^N \epsilon_i\ =\ -2\sqrt2
        \end {equation}

By imposing this constraint by hand,
we can now do the $\phi _ 0$ integration,
which reduces to $\int d \phi _ 0$, which
is divergent but can be factored out.
(If we go to the Minkowski metric,
then the $\phi _ 0$ integral
becomes a Fourier integral, so we have a delta
function which enforces the previous condition.)

The functional integration over $X$ and $\tilde \phi$ is
now trivial,
and in fact reproduces the original Shapiro-Virasoro amplitude,
except that the energy factors are shifted by the various
constraints on energy.
The shifted Shapiro-Virasoro amplitude for the
scattering of $N + 1$ tachyons, in turn, can be written as:

\begin {equation}
A_{\chi_1, \dots, \chi_N} (p_1, \dots , p_N)\ =\ \int
\prod_{i=1}^{N-3} d^2 z_i\, \prod _{i<j} \Big |  z_i - z_j \Big | ^{2 q_i
\cdot q_j}
\end {equation}
which is the familiar amplitude, except that
$q_i \cdot q_j = p_i p_j - \epsilon_i \epsilon_j$ and
$\epsilon_i = - \sqrt2 + \chi_i p_i$.

These integrals can be evaluated, yielding:
        \begin {equation}
        A_{+ \dots +-}^{(N,1)} (p_1, \dots , p_N, p_{N+1})\ =\ {\pi^{N-2}
        \over (N-2)! } \prod_{i=1}^N {\Gamma(1 -\sqrt2 p_i )\over
        \Gamma(\sqrt2 p_i)}
        \end {equation}

        The sum rules $\sum_i p_i =0$ and $\sum_i
        \epsilon_i = -2\sqrt2$ must be satisfied. This, in turn,
fixes the value of $p_{N+1}$ to be:

        \begin {equation}
        \sum_{i=1}^N p_i \ =\ - p_{N+1}\ =\ \frac1{\sqrt2} (N-1)\ .
        \end {equation}

        The amplitudes of type $(1, N)$ are obtained by a
        parity flip.

        These amplitudes exhibit a remarkable structure. They
vanish unless there is exactly one particle with $\chi = -1$,
with all other particles having positive chirality.
This amplitude is also unusual because of the presence of
poles at $p _ i = n / \sqrt 2$ on each external leg, which
does not appear in ordinary field theory.
These \lq\lq leg poles," which occur only at
discrete values of momenta, correspond to the discrete states.
(These leg poles, from a field theory point of view, can be
dropped if we rescale our operators.)

Unfortunately, we have glossed over several
difficult questions in hastily arriving at this amplitude.
The problem with this amplitude
is that we have taken a vanishing
cosmological constant $\mu = 0$,
as well as imposing by hand an additional
constraint on the momenta.

It is possible to remove these constraints
by using a clever trick, which is to
\lq\lq analytically continue in integers."$^{23}$
This may sound mathematically strange, but
can probably be rigorously justified.
To see how this is done, let us first perform
the $\phi _ 0$ integration without
imposing these constraints:

\begin {equation}
\Big \langle \prod _ i T _ { \chi _i } \Big \rangle =
\mu ^ s \Gamma ( -s )
\Big \langle \prod _ i T _ { \chi _ i } \left ( \int d ^ 2 z
e ^ { - { \sqrt 2 } \tilde \phi } \right ) ^ s
\Big \rangle
\end {equation}
where we have used the integral:

\begin {equation}
\mu ^ s \Gamma ( -s ) =
\int _ 0 ^ \infty d A \, A ^ { -s -1 }
e ^ { - \mu A }
\end {equation}
with the definitions
$A = e ^ { - { \sqrt 2 } \phi _ 0 }$ and:

\begin {equation}
- {\sqrt 2 } s = \sum _ i \epsilon _ i + Q
\end {equation}

In general, the matrix element over
the tachyon vertex functions can easily be
calculated for positive integers $s$, but is
problematic for non-positive integers.
Let us, for example, take the case of
four-tachyon scattering, where $k _ 1, k _ 2 , k _ 3$
are positive and $k _ 4$ is negative. Then the
scattering amplitude, for positive integer $s$, can
be evaluated as in the shifted Shapiro-Virasoro
amplitude. The matrix element becomes:

\begin {eqnarray}
A ( k _ i ) &\sim&
\mu ^ s \Gamma ( -s )
\int d ^ 2 z \, | z | ^  { 2 ( k _ 1 k _ 4 - \epsilon _ 1 \epsilon _ 4 )}
| 1 - z | ^ { 2 ( k _ 3 k _ 4 -\epsilon _ 3 \epsilon _ 4 ) }
\nonumber \\
& \times &
\prod _ { i = 1 } ^ s
\int d ^ 2 w _ i
\,
| w _ i | ^ { 2 {\sqrt 2 } \epsilon _ 1 }
| 1 - w _ i | ^ { 2 {\sqrt 2 } \epsilon _ 3 }
| z - w _ i | ^ { 2 s}
\nonumber \\
& \times &
\prod _ { 1 \leq i < j \leq s }
| w _ i - w _ j | ^ { - 4 }
\end {eqnarray}

Evaluating these integrals for positive integer $s$, we have:
\begin {equation}
A ( k _ i ) \sim
( s + 1 ) \mu ^ 2
\Delta ^ s ( 1 )
\prod _ { i = 1 } ^ 4
( - \pi ) \Delta [
{ 1 \over 2 } \epsilon ^ 2 - k _ i ^ 2 ]
\end {equation}
where $\Delta ( x ) = \Gamma ( x ) / \Gamma ( 1-x)$.

Now let us rescale and eliminate the leg poles. We
then have:

\begin {equation}
A (k _ 1 , k _ 2 , k _ 3 , k _ 4 ) \sim
( 1 + s ) \mu ^ s
\end {equation}

In the same way, we can now evaluate the $N$ point scattering
amplitude for arbitrary $\mu$.
For positive integer $s$, we now have:

\begin {equation}
\Big \langle \prod _ { i = 1 } ^ { N + 1 }
T _ { \chi _ i }
\Big \rangle \sim
\left(
{ \partial \over \partial  \mu } \right )
^ { N-2 } \mu ^ { s + N - 2 }
\end {equation}

We now perform the \lq\lq analytic continuation in integers,"
and claim that this formula actually works for arbitrary
$s$, not just positive integers. In this way, we re-derive
the formula found in matrix models in eq. (\ref{eq:mu}).

We should mention that the
expressions for the $N$-point tachyon scattering amplitudes
were calculated using conformal field theory, not string
field theory. We will return to this calculation
of the scattering amplitudes when we
introduce the full non-polynomial string field action.

        \section {Discrete States in Liouville Theory}

        Let us look for discrete states more systematically
        by constructing the physical states within the string
field $| \Phi \rangle$
        using the oscillators of the $X$ and $\phi$ fields.
        Introducing the oscillators $\alpha_n$ and $\beta_n$ through:

        \begin {equation}
        \partial _z X \ =\ -i \sum_n \alpha_n z^{-n-1}\ , \qquad\quad \partial
        _z \phi\
        =\ -i \sum_n \beta_n z^{-n-1}
        \end {equation}
We define:
        \begin {equation}
        \alpha^\mu_n\ =\ (i\beta_n, \alpha_n)\ , \qquad q^\mu\ =\
        (\epsilon,p)\ ,\qquad Q^\mu\ =\ (Q,0) \ .
        \end {equation}

        The Virasoro generators take the form
        \begin {eqnarray}
        L_n\ & =& \ (q^\mu +\frac{n+1}2 Q^\mu) \alpha_{\mu,n} + \sum_k :
        \alpha^\mu_{n+k} \alpha_{\mu, -k} : \qquad \quad n\neq 0
        \nonumber \\
        L_0 \ & =& \ q_\mu ( q^\mu + Q^\mu) + \sum_k : \alpha^\mu_{-k}
        \alpha_{\mu,k} :
        \nonumber \\
        \end {eqnarray}
        where $[\alpha^\nu_n , \alpha^\mu_m] = n\,\delta_{n+m,0}\,
        \eta^{\mu\nu}$, $\eta_{\mu\nu} = {\rm diag}(-1,1)$ and the
        indices are raised and lowered with the $\eta$ metric.

Now let us construct some of the higher excited states
which are solutions of the gauge constraints. The
first excited state (for open strings) is given by:

        \begin {equation}
        | \psi _ L \rangle \ = \epsilon_{\mu}\, \alpha^\mu_{-1}\,
        | q \rangle \ .
        \end {equation}

        We can always generalize this to the closed
string case by doubling the number of states, so that we use
$\epsilon _ { \mu \nu } = \epsilon _ \mu \epsilon _ \nu$.
        The Virasoro conditions now give
        \begin {equation}
        (q_\mu + Q_\mu)\, \epsilon ^\mu\ =\ 0 \ , \qquad\quad
        (q_\mu + Q_\mu)\, q^\mu\ =\ 0 \ .
        \end {equation}

        Notice also that the state $q_\mu \alpha^\mu_{-1}
        | q \rangle $ (with
        polarization $\epsilon ^\mu = q^\mu$) is a pure gauge state, $L_{-1}
        | q \rangle $.

        For general $q_\mu$, the unique solution of the Virasoro conditions
        is $\epsilon _\mu \sim q_\mu$, which corresponds to the pure gauge
state. This
        is in accord with the naive light-cone argument that there are no
        physical oscillator states in two-dimensional string theory.

There are two important exceptional momenta, however.

        First, we can set        $q^\mu = 0$.  Then the gauge
        symmetry becomes trivial,
        and the polarization $\epsilon _\mu=(0,1)$
yields a non-trivial
        physical state, whose vertex operator is
        $\partial  X\overline \partial  X$.}

       Second, we can choose $ q^\mu =- Q^\mu$.  In this case the constraints
        are trivially satisfied, and the polarization $\epsilon _\mu = (0,1)$
again
        gives a physical state. The corresponding operator is $\partial
        X \overline\partial  X e^{-2\sqrt2 \phi}$.

The reason why Liouville theory possesses these strange
discrete states, as we mentioned earlier, is because
the Liouville action is not translationally invariant
in $\phi$.
When imposing the light-cone gauge, we find that we cannot
eliminate all higher string degrees of freedom. Thus, the
existence of these discrete states is a direct consequence of
the breakdown of translation invariance.

In this way, we can show that these discrete states persist for
all higher levels. At first, it may seem to be an impossibly tedious
task to explicitly calculate operator expressions for these
higher discrete states. However, using the representations
of the conformal group, this is actually not a difficult task.

To do this, we must examine representations of the $c =1$ algebra
created by compactifying on a circle. It is known that
we can construct generators of the $SU(2)$ algebra
defined on the circle, given by $\partial X$ and $e ^ { \pm i X}$.
These operators, in turn, can be written as Kac-Moody operators
$H _ 3$ and ladder operators $H _ \pm$:

        \begin {equation}
        H_\pm(z)\ =\ \oint {du\over 2\pi i} : e^{\pm i X(u+z)}:\ , \qquad\quad
        H_3(z)\ =\ \oint {du\over 4\pi} \partial  X(u+z)\ .
        \end {equation}

        The allowed values of the chiral (rescaled)
momenta are $p=n/2$,
        and the simplest primary fields are the
        $SU(2)$ highest weight states:
        \begin {equation}
        \psi_{J,J}(z)\ =\ e^{iJX}(z)
        \end {equation}
        where $J=0, 1/2, 1, 3/2, \ldots$.

Given the fact that $\phi _ { J,J}$ transforms as the highest
weight state, we can compute the other members of the representation
by hitting this state with lowering ladder operators
$H _ -$.  By repeatedly acting with the
lowering operator $H_-$, one builds up
        a spin-$J$ $SU(2)$ multiplet of primary fields $\psi_{J,m}$, $m\in \{J,
        J-1, \cdots, -J\}$, with conformal dimension $\Delta = J^2$:

        \begin {equation}
        \psi_{J,m}(z)\ = \left [ {(J+m)!\over (J-m)! (2J)!}\right ]^{1/2}
        \left[\oint {du\over 2\pi i} : e^{-iX(z+u)}:
        \right]^{J-m} \, \psi_{J,J}(z)\ .
        \end {equation}

So far, these states are only functions of the string variable $X$.
In order to couple it to two-dimensional gravity, we must
\lq\lq dress" the state by inserting a vertex operator
for the $\phi$ field, in order to obtain
physical operators with dimension one:

        \begin {equation}
        \Psi_{J,m}^\pm (z) \ =
        \left [ (J+m)! (J-m)! (2J)!\right ]^{1/2}
        \psi_{J,m}(z)\, :
        e^{\epsilon^\pm_J \phi}(z) :
        \end {equation}
        where $\epsilon^\pm_J\ =\ -1 \pm J$.

One can verify that the operators satisfy
        the Virasoro conditions.
These states are not pure gauge states,
because the corresponding states
        have non-vanishing norms.

Now that we have explicitly written down operator expressions for
the infinite number of discrete states, we will examine the
algebra created by these discrete states.

        \section {$w (\infty)$ in Liouville Theory}

        An important property of the chiral vertex operators
        $\Psi^\pm_{J,m}$ is that they form an interesting
        algebra under the O.P.E.:$^{24}$

        \begin {equation}
        \Psi^+_{J_1,m_1}(z)\, \Psi^+_{J_2,m_2}(0)\ =\ldots+\, {2\over z}
        (J_1m_2 - J_2 m_1)\, \Psi^+_{J_1+J_2-1, m_1+m_2}(0)\, +\, \dots \ ,
        \end {equation}
        where we have shown the only physical operator appearing on the
        right-hand side.
        This is isomorphic to a wedge sub-algebra of $w(\infty)$.

So far, we have only analyzed the states within $| \Phi \rangle$
which are ghost-free. Now let us analyze the states
with ghosts in them and arrange them according to BRST
cohomologies.
    We recall that the physical states of string theory
are given by the cohomology classes of the BRST operator at
ghost number one.
This means that the physical operators are
the ghost number one vertex operators
        which commute with $Q_{BRST}$, modulo
        commutators involving the $Q_{BRST}$ with other operators.
In the BRST formalism,
        the tachyon vertex
        operators (for open strings) is created by multiplying
the usual vertex operator by $c$:

        \begin {equation}
        V^\pm_k (z) \ =\ c(z)\, T _ \pm ( z )
        \end {equation}
        and the discrete vertex operators are given by:
        \begin {equation}
        Y^\pm_{J,m}(z)\ =\ c(z)\, \Psi^\pm_{J+1,m}(z)\ .
        \end {equation}

These operators are non-trivial only at the discrete values of
        the momenta. It can be shown that each $Y$
operator has a partner of adjacent ghost number.
For example, the
$Y^+_{J,m}$ operators have ``partners" at ghost number zero,
        while the $Y^-_{J,m}$ have ``partners" at ghost number two.

Now let us examine the fusion rules
for these ghost
        number zero operators, usually denoted by ${\cal O}_{J,m}$.
Modulo BRST commutators, the fusion rules for these operators form
what is called the \lq\lq ground ring:"$^{25}$
        \begin {equation} {\cal O}_{J,m}{\cal O}_{J',m'}={\cal O}_{J+J',m+m'}
        \end {equation}

Explicitly, their form is given, for the lowest states, by:

        \begin {eqnarray}
        {\cal O}_{0,0}\ & =& \ 1
        \nonumber \\
        {\cal O}_{ {1\over 2},{1\over 2}}\ & =& \ (cb +
        { i \over 2 }  \partial  X
        -{1 \over 2 }  \partial \phi )\, e^{{1 \over 2 }  (iX +\phi)}
        \nonumber \\
        {\cal O}_{ {1\over 2},-{1\over 2}}\
        & =& \ (cb -{ i \over 2 }  \partial X
        -{1 \over 2 } \partial \phi )\, e^{{1 \over 2 }  (-iX +\phi)}
        \nonumber \\
        \end {eqnarray}

The entire ring, via the fusion rules,
can be generated by the last two operators.
One can think of the ring as
        a set of analogues of the identity operator, which occur at the
        discrete momenta.

For open strings, the BRST cohomology is complicated
by the fact that
there exists a non-trivial operator $a$ which is also a BRST commutator:

\begin {equation}
a = [ Q , \phi ] = c \partial \phi + 2 \partial c
\end {equation}
(Normally, this would imply that $a$, a BRST commutator, is also
BRST trivial. Therefore, it would be
uninteresting. However, $\phi$ does not transform as a true
conformal field, so $a$ is actually not BRST trivial, although
it is
BRST invariant. $a$ is therefore BRST non-trivial.)
Thus, multiplication by $a$ increases the ghost
number by one and helps to generate a new BRST cohomology.

To display this cohomology,$^{26}$ it will be useful to use the
language of string field theory.
For open string field theory, we can fix the
gauge by choosing $b _ 0 \Phi = 0$.
The states which satisfy both $Q \Phi = 0$ and
$ b _ 0 \Phi = 0$ are grouped into the
\lq\lq relative cohomology" and are given by
$ {\cal O} _ { J, m}$ and $ Y _ { J,m} ^ +$.
If we relax the $b _ 0 \Phi = 0$ constraint,
then we have a larger set of states, called
the \lq\lq absolute cohomology," given by:

\begin {eqnarray}
G & = & 0 :\quad {\cal O}_{J,m} \nonumber \\
G &  =&  1 :\quad Y^+_{J,m} , \ a {\cal O}_{J,m}\nonumber \\
G & = & 2 :\quad a Y^+_{J,m}\ .
\end {eqnarray}
which exhausts the BRST cohomology for $\epsilon > -1$.

Let ${\cal P} _ { J,m}$ represent the conjugate to
$ Y  _ { J,m}
 ^ +$.
Then there exists another cohomology set, similar to the above,
for $\epsilon < -1$, given by:

\begin {eqnarray}
G & =& 1 :\quad Y^-_{J,m}
\nonumber \\
G & =&  2 :\quad {\cal P}_{J,m} , \ a Y^-_{J,m}
\nonumber \\
G & =&  3 :\quad a {\cal P}_{J,m}\ .
\end {eqnarray}
This completes the open string BRST cohomology.

For closed strings, the situation is a bit more complicated.
We can impose the condition $ ( b _ 0 - \bar b _ 0) \Phi  = 0$
as well as the $b _ 0$ and $\bar b _ 0$.

Thus, there are three sets of cohomologies.
There is the relative cohomology, which satisfies
the $ (L _ 0 - \bar L _ 0 ) \Phi = 0$ constraint
as well as the
$ b _ 0 $,  $\bar b _ 0$, and $ b _ 0 - \bar b _ 0$
constraints.
Then there is the semi-relative cohomology,
which satisfies the $ b _ 0 - \bar b _ 0$ constraint
but not the $ b _ 0$ and $ \bar b _ 0$ constraint.
The semi-relative cohomology is given explicitly by:$^{26}$

\begin {eqnarray}
G & = & 0 :\quad {\cal O}_{J,m}\overline{\cal O}_{J,m'}
\nonumber \\
G & = & 1 :\quad Y^+_{J,m} \overline{\cal O}_{J,m'}, \ {\cal O}_{J,m}
\overline{Y}^+_{J,m'},\ (a +\overline{a})\cdot ({\cal
O}_{J,m}\overline{\cal O}_{J,m'})
\nonumber \\
G & =& 2 :\quad Y^+_{J,m} \overline{Y}^+_{J,m'}, \ (a +\overline{a})\cdot
(Y^+_{J,m} \overline{\cal O}_{J,m'}), \ (a +\overline{a})\cdot
({\cal O}_{J,m} \overline{Y}^+_{J,m'})
\nonumber \\
G & =& 3 :\quad (a +\overline{a})\cdot (Y^+_{J,m}
\overline{Y}^+_{J,m'})\ .
\end {eqnarray}
for $\epsilon > -1 $.

There is another BRST cohomology
set for $\epsilon < -1$, given by:

\begin {eqnarray}
G & = & 2 :\quad Y^-_{J,m} \overline{Y}^-_{J,m'}
\nonumber \\
G & = & 3 :\quad Y^-_{J,m} \overline{\cal P}_{J,m'}, \ {\cal P}_{J,m}
\overline{Y}^-_{J,m'},\ (a +\overline{a})\cdot (Y^-_{J,m}
\overline{Y}^-_{J,m'})
\nonumber \\
 G & = & 4 :\quad {\cal P}_{J,m}\overline{\cal P}_{J,m'},\
(a +\overline{a})\cdot (Y^-_{J,m} \overline{\cal P}_{J,m'}),
\ (a +\overline{a})\cdot({\cal P}_{J,m} \overline{Y}^-_{J,m'})
\nonumber \\
G & = & 5 :\quad (a +\overline{a})\cdot ({\cal P}_{J,m}
\overline{\cal P}_{J,m'}) \ .
\end {eqnarray}

And lastly, there is the absolute cohomology, which satisfies
none of the $b$ constraints.

        \section {Non-polynomial String Field Theory}

Now that we have discussed the preliminaries of the first quantized
Liouville theory, we would like to reanalyze this information within
the context of a second quantized field theory.

One advantage of introducing the string field $\Phi$ is that
we can now assemble the tachyon, discrete states, and higher
BRST trivial states into one field.

        Symbolically, we have:
        \begin {equation}
        | \Phi ( X , b , c , \phi ) \rangle =
        | {\rm tachyon}\rangle +
        | {\rm discrete} \, {\rm states} \rangle
        +
        | {\rm BRST} \, {\rm trivial} \, {\rm states} \rangle
        \end {equation}

Another advantage of introducing string field theory is that
the mysterious $w(\infty)$ symmetry emerges in much the same
way that symmetries emerge in ordinary gauge theory. For example,
in gauge theory, the structure constants of $SU(N)$ emerge as
the coupling of three gauge fields $A _ \mu ^ a$. Similarly,
we can show that the coupling of three string fields
is proportional to the structure constants of $w(\infty)$.
Thus, $w(\infty)$ is a subalgebra of the full gauge
algebra of string field theory.

If $|jm \rangle$ represents a discrete state, then
one can show that the
three-string vertex function of string field theory, taken
on discrete states, yields the $w(\infty)$ structure constants:

        \begin {eqnarray}
        \langle j _ 1 , m _ 1  | \langle j _ 2 , m _ 2 |
        \langle j _ 3 , m _ 3 | V _ 3 \rangle
        & \sim &
        \Big \langle \Psi  _ { j _ 1 , m_ 1 } ( 0 )
        \Psi  _ { j _ 2 , m _ 2 } ( 1 )
        \Psi _ { j _ 3 , m _ 3 } ( \infty) \Big \rangle
        \nonumber \\
        & \sim & ( j _ 1 m _ 2 - j _ 2 m _ 1 ) \delta _ { j _ 3 ,
        j _ 1 + j _ 2 - 1 } \delta _ { m _ 3 ,
        m _ 1 + m _ 2 }
        \end {eqnarray}

Now, let us discuss the action
for $D = 2$ non-polynomial string field theory.
The basic structure of the $D=2$ action must be
identical to the action for the 26 dimensional case.
This is because
the triangulation of moduli space with cylinders of equal
circumference (but arbitrary length) is independent of the
dimension of space-time. The action is therefore:

        \begin {equation}
        {\cal L} = \langle \Phi | \hat Q | \Phi \rangle +
        \sum _ { n = 3 } ^ \infty \alpha _ { n}
        \langle \Phi ^ n \rangle
        \end {equation}
        where $\hat Q = Q _ 0 ( b _ 0 - \bar b _ 0 )$, $Q _ 0$
        is the usual BRST operator, and $\langle \Phi ^ n \rangle$
represents a $n$ string vertex function, such that the $n$ strings
meet to form the topology of a polyhedra. Note that there are
more than one distinct polyhedra at each level.
        For example, there are 2 polyhedra at $N=6$,
        5 polyhedra at $N=7$,
        and 14 polyhedra at $N=8$.

The fact that closed string field theory (defined with strings of
equal string length) requires an additional four-string graph to
reproduce the Shapiro-Virasoro amplitude was first pointed out  in ref.
10, where it was shown that the \lq\lq missing region" of moduli space
could be filled exactly by a four-string diagram with the
topology of a tetrahedron.

For example, let $a _ { ij }$ represent the distance that the
$i$th and $j$th string share in common. Then $a _ { ij } = 0$
if they share no common boundary. Then we have
$\sum _ i a _ { ij } = 2 \pi$ for fixed $j$,
which simply says that the total
circumference of the $j$th closed string is $2 \pi$.

The lengths of the sides of the tetrahedon is therefore
governed by four constraints, representing the four faces
of a tetrahedron:

\begin {eqnarray}
a _ { 12 } + a _ { 13 } + a _ { 14 } & = & 2 \pi
\nonumber \\
a _ { 21 } + a _ { 23 } + a _ { 24 } & = & 2 \pi
\nonumber \\
a _ { 31 } + a _ { 32 } + a _ { 34 } & = & 2 \pi
\nonumber \\
a _ { 41 } + a _ { 42 } + a _ { 43 } & = & 2 \pi
\end{eqnarray}

There are six unknowns $a _ { ij }$ for a tetrahedron.
There are four constraints on them, given above.
So we have a net number of two degrees of freedom, which is
the correct number of moduli
needed to parametrize the Shapiro-Virasoro amplitude.

Similarly, for an $N$ sided polyhedra,
the number of edges or sides is $ 3N-6 $, and the
number of constraints is $N$ for $N$ faces, so that the total
number of degrees of freedom is $2N - 6$,
which is the correct number of Koba-Nielsen variables or moduli
necessary to describe the $N$ string amplitude.
In other words:

\begin {equation}
3N-6 \, {\rm Edges} -
N \, {\rm Faces} =
 2N-6
\, {\rm Moduli}
\end {equation}

For the tetrahedron graph, we can choose $a _ { 12}$ and
$a _ { 13}$ as independent variables. Then solving the
constraints, we find that the missing region is parametrized
by:

\begin {equation}
{\rm
Missing \, region } \, = \,
\left \{ \begin {array} {lll}
a _ { 12 } , a _ { 13 } \leq \pi
\\
a _ { 12 } + a _ { 13 } \geq \pi
\\
        \end {array}
        \right .
        \end {equation}

This missing region is filled precisely by the four-string interaction
with the topology of a tetrahedron.
Similarly, it is not hard to generalize this counting for higher
polyhedra.

The non-polynomial action, in turn, is invariant
under the following gauge transformation:

        \begin {equation}
        \delta | \Phi \rangle =
        | Q \Lambda \rangle +
        \sum _ { n = 1 } ^ \infty \beta _ { n }  | \Phi ^ n \Lambda \rangle
        \end {equation}

        If we insert $\delta | \Phi \rangle$ into the action, we find that
        the result does not vanish, unless:

        \begin {equation}
        ( -1 ) ^ n \langle \Phi || Q \Lambda \rangle +
        n \langle Q \Phi || \Phi ^ { n - 1 } \Lambda \rangle
        + \sum _ { p = 1 } ^ { n-2 }
        C _ p ^ n \langle \Phi ^ { n-p} ||
        \Phi ^ p \Lambda \rangle = 0
        \end {equation}
        where the double bars mean that when we join two polyhedra,
        the common
        boundary has circumference
        $2 \pi$.

(We should point out that the gauge transformation outlined above
is actually anomalous, i.e. the measure of integration
${\cal D} \Phi$ is not invariant under this gauge transformation.
This means that the transition from the classical action to the
quantum one requires additional non-polynomial terms at the
higher loop level. There are two ways in which one can
calculate these higher loop corrections. First, one can
use the Fujikawa method and calculate a recursion relation
which generates the complete quantum action.$^{27}$
Or, one can use BV quantization.$^{28}$)

The problem facing us, however, is that the vertices are anomalous
unless we take into account the proper insertion factors.
The Liouville action possesses terms like $R \phi$, which introduce
curvature singularities on the world-sheet where the strings join.
Thus, we must insert the proper insertion factors at these
singular points in order to maintain gauge invariance.

        To do this, we will find it convenient to bosonize
the $b,c$ ghosts, such that $c = e^ \sigma$ and $b = e ^ { - \sigma}$.
This will allow us to treat $X$, $\phi$ and $\sigma$
on almost the same footing. Then let us introduce the notation:

        \begin {eqnarray}
        Q ^ M & = & \{ 0 , Q , -3 \}
        \nonumber \\
        \phi ^ M & = & \{ X ^ i , \phi , \sigma \}
        \end {eqnarray}

        The insertion factor, placed at the points where strings join,
can now be expressed as (where we adopt a slightly different normalization
for the curvature, to conform with the literature):

        \begin {equation}
        \prod _ { j = 1 } ^ { 2(N-2)}
        \left( e ^ { -  Q ^ M \phi ^ M / 2 } \right) _ j
        \end {equation}
        where $j$ labels the $2(N-2)$ sites where we have curvature
        singularities on the string world sheet.

The $N$-string vertex function $| V _ N \rangle$ can now be written
down as $| V _ N \rangle = \int B _ N |  V _ N \rangle _ 0 $,
where $B _ N$ contains line integrals over $b$ fields described by
Beltrami differentials. The $|V _ N \rangle _ 0$ is the usual
vertex function, given by a series of delta functions representing
the overlap of $N$ strings.
Written out explicitly, $|V _ N \rangle _ 0$ is:$^{9}$

        \begin {eqnarray}
        |V _ N \rangle _ 0 & = &
        \left (
        \prod _ { j = 1 } ^ { 2 ( N-2) } e ^ { - ( Q ^ M \phi ^ M / 2 ) _ j }
\right )
         \int
        \delta ( \sum _ { i = 1 } ^ N
        p _ i ^ M + Q ^ M ) \prod _ { i = 1 } ^ N P_ i
        \nonumber \\
        & \times &
        {\rm exp } \,
        \Bigg \{
        \sum _  { r,s } ^ N \sum _ { n,m = 0}
         ^ \infty
        { 1 \over 2 }
        N _ { nm } ^ { rs } \alpha  _ { -n } ^ { M r }
        \alpha _ { -m } ^ { M  s }
        \Bigg \}
        \nonumber \\
        & \times &
        {\rm exp } \,
        \Bigg \{
        \sum _  { r,s } ^ N \sum _ { n,m = 0}
         ^ \infty
        { 1 \over 2 }
        N _ { nm } ^ { rs } \tilde \alpha  _ { -n } ^ { M r }
        \tilde \alpha _ { -m } ^ { M  s }
        \Bigg \}
        \left ( \prod _ { i = 1 } ^ N d ^ M p _ i | p _ i ^ M \rangle
        \right)
        \end{eqnarray}
        where $P _ i $ represents the operator which rotates the string
        field by $2 \pi$,
        where $j$ labels the insertion points, where we have deliberately
        dropped an uninteresting constant,
        and where the
        state vector $| p _ i ^M \rangle $ and
        the Neumann functions, which describe the world-sheet, are defined in
the
usual way.

Naively, we expect that
\begin {equation}
\sum _ { i = 1 } ^ N Q _ i | V _ N \rangle _ 0 = 0
\end {equation}

The proof of this statement is far from obvious.
The calculation of the
anomalous term is quite non-trivial, involving subtle
point-splitting methods at the point where strings
join.

        \section {Proof of Gauge Invariance}

        The proof of gauge invariance is complicated by the
        fact that there are subtle but important anomalies
        in the calculation of BRST invariance.
        For three strings, the calculation is performed by
writing the BRST operator as a line integral over the three
strings:
        \begin {eqnarray}
        \sum _ { i = 1} ^ 3 Q _ i | V _ 3 \rangle _ 0  & = &
        \oint _ { C _ 1 + C _ 2 + C _ 3 }
        { d \rho \over 2\pi } c ( \rho )
        \nonumber\\
        & \times &
        \left \{
        - { 1 \over 2 }
        ( \partial _ z \phi ^ \mu ) ^ 2
        + { d c \over d\rho } b ( \rho )
        + { Q \over 2 } ( \partial _ z \phi ^ \mu ) ^ 2
        \right \} | V _ 3 \rangle _ 0
        \end {eqnarray}
        where $C _ i$ are infinitesimal curves which
        together comprise circles which go around
        $\rho ( z _ i)$ and $\rho (\tilde z _ i )$.

        The calculation is rather tedious, so we will only quote
the final result, which is given by$^{9}$:

        \begin {equation}
        \left \{  p c ( z _ 0 ) \left [
        { D \over 24 } - { 13 \over 12 } + { 1 \over 24 }
        + { 1 \over 8 } Q ^ 2 \right ]
        + { d c ( z _ 0 ) \over d z }
        \left [ { 5 D \over  96 }
        - { 65 \over 48 }
        + { 5 \over 96 }
        + { 5 \over 32 } Q ^ 2 \right ]
        \right \} | V _ 3 \rangle _ 0
        \end {equation}
        which cancels if:

        \begin {equation}
        D - 26 + 1 + 3 Q ^ 2 = 0
        \end {equation}
        which is precisely the consistency equation for Liouville theory
        in $D$ dimensions.
        To check the accuracy of this result, we can also
        perform the calculation using bosonized co-ordinates,
        which yields:
        \begin {equation}
        \left \{ p e ^ { \sigma ( z _ 0 ) } \left [
        { D + 2 \over 24 }
        + { 1 \over 8 } ( Q ^ 2 - 3 ^ 2  ) \right ]
        + { d e ^ { \sigma ( z _ 0 ) } \over d z }
        \left [ { 5 ( D + 2 ) \over  96 }
        + { 5 \over 32 } ( Q ^ 2 - 3 ^ 2 ) \right ]
        \right \} | V _ 3 \rangle _ 0
        \end {equation}
        Once again, we find that the anomaly cancels if we set:

        \begin {equation}
        D + 2 + 3 ( Q ^ 2 - 3 ^ 2 )= 0
        \end {equation}
        as desired.
        Since the point-splitting method isolates the anomalous
        contribution of each vertex of the polygon, it is then trivial
        to prove that all $N$-point vertices are also
        BRST covariant.

        \section {Derivation of the Shifted Shapiro-Virasoro Amplitude}

        To show that the non-polynomial string field theory is
        correct, we must also show that it reproduces the
        correlation functions of Liouville theory.
This is non-trivial, because the conformal maps used in
open string field theory cannot be used for the closed string
case.

        To perform this calculation, we first need the conformal
        map between the complex $z$-plane and the
        multi-sheeted $\rho$ plane, which describes the collision
        of several strings of equal perimeter.
This is not easy, because the conformal map used in the
covariant open string field theory calculation
cannot be used here.

By analyzing the zeros and singularities of the conformal map, we
find that it is given
by$^{29}$:
        \begin {equation}
        { d \rho (  z )
        \over d z }
        = C { \prod _ { i = 1 } ^ { N - 2 }
        \sqrt  { ( z - z _ i ) ( z -  \tilde z _ i ) }
        \over
        \prod _ { i = 1 } ^ N  ( z - \gamma _ i ) }
        \end {equation}
        where the $N$ variables
        $\gamma _ i$ map to
        points at infinity (the external
        lines in the $\rho$ plane) and the $N-2$ pair of variables
        $( z _ i, \tilde z _ i ) $ map to the
        points where two strings collide, creating the
        $i$th vertex (which are interior points in
        the $\rho$ plane).

        The number of unknowns in this map
        minus the number of moduli equals the number of constraints:

\begin {equation}
{\rm Unknowns} \, - \, { \rm Moduli } = {\rm Constraints}
\end {equation}

More explicitly, we have:
        \begin {eqnarray}
        \{ C , z _ i , \tilde z _ i , \gamma _ i\} & \rightarrow &
        6N-6 \, {\rm unknowns}
        \nonumber \\
        \{ \rho ' ( \gamma _ i ) , \rho ( z _ i) - \rho ( \tilde z _ i )
        \} & \rightarrow & 4N \, {\rm constraints}
        \nonumber \\
        \{ \tau _ { ij } + i \theta _ { ij } \} & \rightarrow &
        2N - 6 \, {\rm moduli}
        \end {eqnarray}
        where $\tau _ { ij }$ is the distance separating the
$i$th and $j$th vertex functions, and $\theta _ { ij }$ is the
relative angle separating these two vertices.

        For the four point function, the map
        can be integrated using ordinary analytical functions.
        We use the identity:

        \begin {equation}
        { ( z - z _ 1 ) ( z - \tilde z _ 1 )
        ( z - z _ 2 ) ( z - \tilde z _ 2 )
        \over \prod _ {i = 1 } ^ 4 ( z - \gamma _ i ) }
        =
        1 + \sum _ { i = 1 } ^ 4
        { A _ i \over z - \gamma _ i }
        \end {equation}
        where we define $ z _ i = i a _ i + b _ i$ and $\tilde z _ i
        = - i a _ i  + b _ i $ for {\it complex} $a _ i$ and $b _ i$, and:

        \begin {equation}
        A _ i = {  \left [ ( \gamma _ i - b _ 1 ) ^ 2 + a _ 1 ^ 2 \right ]
        \left [ ( \gamma _ i - b _ 2 ) ^ 2 + a _ 2 ^ 2  \right ]
        \over
        \prod _ { j = 1 , j \neq i } ^ 4
        ( \gamma _ i - \gamma _ j ) }
        \end {equation}

        Then we can split the integral into two parts, with the result:

        \begin {eqnarray}
        \rho ( z ) & = &
        \rho _ 1 ( z ) + \rho _ 2 ( z )
        \nonumber \\
        \rho _ 1 ( z ) & = &
        \int _ { y _ 1 } ^ y
         { N dz
        \over \sqrt { ( z - z _ 1 ) (z  - \tilde z _ 1 )
        ( z - z _ 2 ) ( z - \tilde z _ 2 ) } }
        \nonumber \\
        \rho _ 2 ( z ) & = &
        \sum _ { i = 1 } ^ 4
        \int _ { y _ 1 } ^ y
        { N A _ i dz \over
        ( z - \gamma _ i )
        \sqrt { ( z - z _ 1 )( z - \tilde z _ 1 )( z - z _ 2 )
        (z - \tilde z _ 2 ) } }
        \end {eqnarray}

        Written in this form, we can now perform all integrals exactly,
        using third elliptic integrals.
        It is then easy to show:

        \begin {eqnarray}
        \rho _ 1 ( z ) & = &
        N g  F ( \phi , k ' ) =
        N g {\rm tn } ^ { -1 }
        \left [ \tan \phi , k
        ' \right ]
        \nonumber \\
        \rho _ 2 ( z ) & = &
        \sum _ { i = 1 } ^ 4
        { g N A _ i
        \over
        a _ 1 + b _ 1 g _ 1 - g _ 1 \gamma _ i }
        \Bigg (
        g _ 1 F ( \phi , k ' )
        \nonumber \\
        & + &
        { \omega _ i - g _ 1 \over
        1 + \omega _ i ^ 2 }
        \left [ F ( \phi , k ' ) + \omega _ i ^ 2
        \Pi ( \phi , 1 + \omega _ i ^ 2 , k ' )
        +
        \omega _ i ( \omega _ i ^ 2 + 1 ) f _  i
        \right]
        \Bigg )
        \end {eqnarray}

        where:
        \begin {eqnarray}
        \omega _ i &=& { a _ 1 + b _ 1 g _ 1 - \gamma _ i g _ 1
        \over b _ 1 - a _ 1 g _ 1 - \gamma _ i }
        \nonumber \\
        f _ i &=&
        { 1 \over 2 } ( 1 + \omega _ i ^ 2 ) ^ { - 1 /2 }
        ( k ^ 2 + \omega _ i ^ 2 ) ^ { - 1/2 }
        \nonumber \\
        && \times \ln
        { ( k ^ 2 + \omega _ i ^ 2 ) ^ { 1/2 }
        - ( 1 - \omega _ i ^ 2 ) ^ { 1/2 }
        { \rm dn } u
        \over
        ( k ^ 2 + \omega _ i ^ 2 ) ^ { 1/2 }
        + ( 1 + \omega _ i ^ 2 ) ^ { 1/2 }
        { \rm dn } u }
        \nonumber \\
        \phi & = & {\rm arc tan}
        \left( { y - b _ 1 + a _ 1 g _ 1
        \over a _ 1 + g _ 1 b _ 1 - g _ 1 y } \right )
        \end {eqnarray}
        and where:
        \begin {eqnarray}
        A ^ 2 & = & ( b _ 1 + b _ 2 ) ^ 2 +
        ( a _ 1 + a _ 2 ) ^ 2
        , \, \,  B ^2  =
        ( b _ 1 - b _ 2 ) ^ 2 + ( a _ 1 - a _ 2 ) ^ 2
        \nonumber \\
        g _ 1 ^ 2 &= & [4 a _ 1 ^ 2 - ( A - B) ^ 2 ] /
        [ ( A + B ) ^ 2 - 4 a _ 1 ^ 2 ]
        , \, \,
        g  =  2 / ( A + B )
        \nonumber \\
        y _ 1 & = & b _ 1 - a _ 1 g _ 1
        , \, \, { k '} ^ 2 =  1 - k ^ 2 = 4AB / ( A + B ) ^ 2
        \nonumber \\
        u &=  & {\rm  dn} ^ { -1 } ( 1 - { k '}  ^ 2  {\rm sin} ^ 2 \phi )
        \end {eqnarray}

        After a considerable amount of algebra, this
expression simplifies
        to:$^{29}$

        \begin {eqnarray}
        \rho ( z ) & = &
        \sum _ { i = 1 } ^ 4
        { g N A _ i \over
        a _ 1 + b _ 1 g _ 1 - g _ 1 \gamma _ i }
        { \omega _ i - g _ 1 \over 1 + \omega _ i ^ 2 }
        \nonumber \\
        && \times
        \left [ \omega _ i ^ 2
        \Pi ( \phi , 1 + \omega _ i ^ 2  , k ' )
        + \omega _ i  ( \omega _ i ^ 2 + 1 )
         f _i \right ]
        \end {eqnarray}

        Now that we have the explicit conformal map for four-string
        scattering, it is straightforward to find the Jacobian which
        takes us from the moduli describing string scattering to the
        Koba-Nielsen variables.
        The string scattering is described by $\hat \tau =
        \tau + i \theta$, where $\tau$ is the length of the
        intermediate string, and $\theta$ is the relative angle
        that strings one and two are rotated with respect to
        string three and four.

        Similarly, with a bit of work
one can calculate expressions for $d \hat \tau$
and $d \hat x$. Putting everything together, we find:

        \begin {equation}
        { d \hat \tau \over d \hat x }
        =
        - { \pi N \over
        2 K( k ) g \hat x ( 1 - \hat x )
        ( \gamma _ 1 - \gamma _ 3 )( \gamma _ 2 - \gamma _ 4)}
        \end {equation}

        If we take only the tachyon component of
        $|\Phi \rangle$, then the
        four point amplitude can be written as:

        \begin {eqnarray}
        A _ 4 &=& \langle V _ 3 | { b _ 0 \bar b _ 0 \over
        L _ 0 + \bar L _ 0 - 2 } | V _ 3 \rangle
        \nonumber \\
        & = &
        \int d \tau d \theta
        \, \Big \langle
        V ( \infty ) V ( 1 )
        \left ( \int _ C d z { dz \over dw } b _ { zz } \right)
        \left ( \int _ C { d \bar z \over d \bar w } b_ { \bar z \bar z }
        \right )
        V ( \hat x ) V ( 0 ) \Big \rangle
        \nonumber \\
        & = &
        \int d ^ 2 \hat \tau
        \, \Bigg | {\rm exp } \, \left [
        \sum _ { i } \left ( i p _ i \cdot
        \phi ( i ) + \epsilon _ i  \phi ( i ) \right ) \right )
        A _ G \Bigg | ^ 2
        \end {eqnarray}
        where we must sum over all permutations so that
        we integrate over the entire complex plane,
        where $b _ 0$ defined in the $\rho$ plane
        transforms into $\int _ C dz (dz/dw) b _ { zz }$ in the
        $z$-plane,
        where $C$ is the image in the $z$-plane of a
        circle in the $\rho$ plane
        which slices the intermediate closed string,
        where $V ( z ) = c ( z ) \tilde c ( z ) V _ 0 ( z ) $,
        where $V _ 0$ is the tachyon vertex without ghosts,
        and
        where the ghost part $A _ G$ equals:

        \begin {eqnarray}
        A _ G &=&
        \int _ C { d z \over 2 \pi i }
        { d z \over d w }
        {\rm exp }
        \left \{
        - \sum _ { i \leq j }
        \langle \sigma _ i \sigma _ j
        \rangle +
        \sum _ j
        \langle \sigma _ j \sigma _ + ( z ) \rangle
        \right \}
        \nonumber \\
        & = &
        \int _ C { d z \over 2 \pi i }
        { d z \over dw }
        { \prod _ { i < j } ( \gamma _ i - \gamma _ j )
        \over \prod _ { j = 1 } ^ 4
        ( z - \gamma _ j ) }
        \noindent \\
        & = &
        2 { g \over \pi c }  \hat x ( 1 - \hat x ) K ( k )
        ( \gamma _ 1 - \gamma _ 3 ) ^ 3 ( \gamma _ 2 - \gamma _ 4 ) ^ 3
        \end {eqnarray}
        (Notice that we have made a conformal transformation from the
        $\rho$ world sheet to the $z$ complex plane. In general, we pick
        up a determinant factor, proportional to the
        determinant of the Laplacian defined on the world sheet.
        However, after making the conformal transformation,
        we find that the determinant of the Laplacian
        on the flat $z$-plane
        reduces to a constant. Thus, we can in general ignore this
        determinant factor.)

        Putting the Jacobian, the ghost integrand, and the
        string integrand together, we finally find:

        \begin {equation}
        A _ 4 = \int d ^ 2 \hat x \Big |
        \hat x ^ { 2 p _ 1 \cdot p _ 2 }
        ( 1 - \hat x ) ^ { 2 p _ 2 \cdot p _ 3 } \Big | ^ 2
        \end {equation}
        In two-dimensions, we have
        $p _ i \cdot p _ j  = p _ i p _ j -
        \epsilon _ i \epsilon _ j$ where
        $\epsilon _ i = - \sqrt {2} + \chi _ i p _ i $,
        where $\chi$ is the \lq\lq chirality" of the tachyon
        state,
        so we reproduce the integral found in matrix models
        and Liouville theory. (The amplitude is non-zero only if
        the chiralities are all the same except for one external line.)

We must say, however, that our results are only
good for $\mu = 0$. We saw earlier that,
by \lq\lq analytically continuing in integers,"
we could use conformal field theory and
manipulate the shifted
Shapiro-Virasoro amplitude, thereby formally
deriving the matrix model
result.
Unfortunately, we cannot use a similar trick for the
non-polynomial calculation.
Because of the highly non-linear nature of the
$\mu \neq 0$ Liouville theory,
it is not known whether we can derive the
$\mu \neq 0 $ amplitudes using
some generalization of the $\mu = 0$ non-polynomial
string field theory.
This is an open question.

        \section {Conclusion}

In summary, we have been able to formulate two-dimensional
string field theory in at least four different ways.
In each method, we have a different way of viewing
the origin of the
discrete states and the $w(\infty)$.

The first three methods (free fermion theory, collective
field theory, and temporal gauge field theory)
are easy to work with, because
all gauge degrees of freedom have been explicitly eliminated.
However, this obscures the string picture underlying the theory.
Thus, discrete states and $w(\infty)$ are easy to display in
this formalism, but rather difficult to explain intuitively.

By contrast, the Liouville string field theory has all degrees of
freedom intact. Calculationally, it is more difficult to use than
the other string field theories. But since the string picture
is left untouched, the discrete states and
$w(\infty)$ appear in much the same way as in ordinary
Yang-Mills theory, as byproducts of the full gauge invariance of
the theory.

In principle, all four field theories should emerge as gauge
fixed versions of the same theory.
However, because of the complicated nature of the interactions
(both cubic and non-polynomial), this is still an open question.

We should also point out that string field theory
still faces many problems. Most of the results of
two-dimensional gravity and strings were first found in the
first quantized formalism, not the second.
String field theory still rarely yields new insights
beyond the perturbative, first quantized approach.
Also,
string field theory
is still
formulated in a background dependent fashion, which is
one reason why it is not clear how to accomodate
black hole solutions in this formalism.
All these problems, we are confident, will be solved with time.

        \nonumsection{Acknowledgements}
        \noindent
We would like to acknowledge partial support from
CUNY-FRAP 6-64435 and NSF PHY-9020495.

        \nonumsection{References}

\end{document}